\documentclass[prd, twocolumn, floatfix, nofootinbib, notitlepage, showkeys, showpacs]{revtex4-1}
\usepackage{amsmath}
\usepackage{amssymb, latexsym, mathrsfs}
\usepackage{bm}
\usepackage{caption}
\usepackage{subcaption} 
\usepackage{color}
\usepackage{graphicx}
\usepackage{dcolumn}
\usepackage{epsfig}
\usepackage{graphicx}
\usepackage{hyperref}
\usepackage{ragged2e}
\DeclareCaptionJustification{justified}{\justifying}
\hypersetup{
    colorlinks=true,
    linkcolor=red,
    citecolor=blue,
} 

\newcommand{\be}{\begin{equation}}
\newcommand{\ee}{\end{equation}}
\newcommand{\bs}{\begin{split}}
\newcommand{\es}{\end{split}}
\newcommand{\bea}{\begin{eqnarray}}
\newcommand{\eea}{\end{eqnarray}}

\begin{document}

\title{Upon the horizon's verge:\\Thermal particle creation between and approaching horizons}

\author{Diego Fernández-Silvestre${}^{1}$}
\email{dfsilvestre@ubu.es}
\author{Michael R. R. Good${}^{2,3}$}
\email{michael.good@nu.edu.kz}
\author{Eric V. Linder${}^{3,4}$}
\email{evlinder@lbl.gov}

\affiliation{${}^1$Departamento de Física, Universidad de Burgos, 09001 Burgos, Spain\\
${}^2$Physics Department, Nazarbayev University, Astana 010000, Qazaqstan\\
${}^3$Energetic Cosmos Laboratory, Nazarbayev University, Astana 010000, Qazaqstan\\
${}^4$Berkeley Center for Cosmological Physics \& Berkeley Lab, 
University of California, Berkeley, CA 94720, USA}

\begin{abstract}
Quantum particle creation from spacetime horizons, or accelerating boundaries in the dynamical Casimir effect, can have an equilibrium, or thermal, distribution. Using an accelerating boundary in flat spacetime (moving mirror), we investigate the production of thermal energy flux despite non-equilibrium accelerations, the evolution between equilibrium states, and the ``interference'' between horizons. In particular, this allows us to give a complete solution to the particle spectrum of the accelerated boundary correspondence with Schwarzschild-de Sitter spacetime.
\end{abstract} 
%04.62.+v (Quantum fields in curved spacetime)
%44.40.+a Thermal radiation
\date{\today} 

\maketitle

%%%%%%%%%%

\section{Introduction} \label{sec:intro} 

Particle creation is a striking phenomenon manifested by quantum field theory in conjunction with gravitation, acceleration, or spacetime horizons. Of particular interest is a thermal spectrum of produced particles, like that associated with Hawking radiation \cite{Hawking:1974sw}, or the Davies-Fulling-Unruh effect \cite{Fulling:1972md,Davies:1974th,unruh76}. Generally, a thermal particle distribution is expected to be accompanied by a constant energy flux \cite{carlitz1987reflections}. This characteristic is usually associated with equilibrium, but this is not the full story. For example, constant acceleration produces no energy flux \cite{Davies:1976hi,Davies:1977yv,Birrell:1982ix}. 

We analyze several situations within the context of the dynamical Casimir effect. An example of interest is the so-called accelerated boundary correspondence (ABC), where an analogy is established between a curved spacetime and an accelerating boundary (moving mirror). The ABC has been explored for several known solutions, including Schwarzschild \cite{Good:2016oey}, Reissner-Nordstr\"om \cite{good2020particle}, and Kerr \cite{Good:2020fjz}, reducing the complexity of particle creation in higher dimensional curved spacetimes to a simple flat (1+1)-dimensional spacetime (e.g.\ see the remarkable simplicity afforded in the non-relativistic regime \cite{Ford:1982ct}) of an accelerating mirror slicing through the vacuum, even in the case of unitary evolution \cite{Reyes:2021npy}, see also e.g.\ the Schwarzschild-Planck trajectory \cite{Good:2019tnf,Moreno-Ruiz:2021qrf,Good:2020fsw}. This approach allows analytic, or 
numerical, calculation of the energy flux and particle spectrum detected by a distant observer, and under what dynamical conditions the energy flux appears constant. 

The classic case of an eternally thermal energy flux is the Carlitz-Willey mirror \cite{carlitz1987reflections,CW2lifetime}, but we investigate how the energy flux and particle spectrum {\it approaches\/} a thermal distribution, especially using a transition between two dynamical states -- whether two asymptotically constant accelerations, two accelerations associated with different temperatures, or a dynamical Casimir effect from motion between two bounding horizons. 

Section~\ref{sec:trscon} derives how to form a transition from asymptotically constant acceleration to nevertheless produce non-zero energy flux at a constant level. In Section~\ref{sec:trscwf} we solve for the particle spectrum between two thermal states 
corresponding to different temperatures. Section~\ref{sec:twohor} generalizes this to two horizons, i.e.\ boundaries in the spacetime, and demonstrates the analysis specifically for the ABC with Schwarzschild-de Sitter spacetime, i.e.\ a black hole embedded in an expanding universe with positive cosmological constant. Analysis of hybrid cases is considered in Section~\ref{sec:df} for connecting null-self-dual (Davies-Fulling) double horizons as well as in Appendix~\ref{sec:trsconcw} for connecting thermal states between constant and Carlitz-Willey accelerations. We emphasize the physical implications and insights, and conclude, in Section~\ref{sec:concl}. 

%%%%%%%%%%

\section{Thermal Flux from Transition between Constant Accelerations} \label{sec:trscon} 

Constant acceleration produces no energy flux. However, an acceleration which is asymptotically constant can produce an energy flux, even a thermal plateau, all the way to the asymptote (see for example Sec.~VI.D of \cite{Good:2021iny}). We derive here the requirements for this, and in later sections connect past or future asymptotic constant regimes in such as way as to give sets of mirrors with thermal eras. First we work in coordinate time $t$, then in the null coordinate label function $f(v)$, appropriate to the computation of the particle spectrum. 

%%%%%%%%%%

\subsection{Transition in Coordinate Time} \label{sec:trst} 

A mirror with trajectory $z(t)$ has coordinate velocity $\dot z(t)=dz/dt$, Lorentz factor $\gamma(t)=(1-\dot z^2)^{-1/2}$, and proper velocity (celerity)
\be 
w(t)\equiv \gamma\dot z=\frac{\dot z}{\sqrt{1-\dot z^2}}\ . 
\ee 
Note that the proper acceleration is $\alpha(t)=\dot w=\gamma^3\ddot z$. We can rewrite this as 
\be 
\dot z(t)=\left(1+w^{-2}\right)^{-1/2}\approx 1-\frac{1}{2}w^{-2}+\frac{3}{8}w^{-4}+\dots 
\label{eq:zwexp} 
\ee 
where the approximation holds for large $w$, appropriate for our study of the asymptotic behavior with velocities near the speed of light. 

Using these expressions, the energy flux is then given as 
\bea 
-12\pi F(t)&=&\frac{(1-\dot z^2)\dddot z+3\dot z\ddot z^2}{(1+\dot z)^2(1-\dot z)^4}\\ 
&\approx&4\frac{w^{-5}\ddot w+w^{-6}\alpha^2\cdot 0 +w^{-8}\alpha^2\cdot 0}{w^{-8}} \label{eq:ddotwmid} \\
&\approx& 4w^3\ddot w\, \label{eq:fzp1} 
\eea 
when $\dot z\to+1$. First, we see that if $\ddot w=\dot\alpha=0$, then the energy flux 
is zero. That is, a uniform constant acceleration gives no energy flux. 

Thermal particle creation gives a constant energy flux, requiring $\ddot w\sim w^{-3}$. Suppose $w\sim t^n$. Then, the condition for thermal flux $\dot F(t)=0$ requires either $n=1/2$ or $n=1$. The $n=1/2$ case corresponds to the well known Carlitz-Willey \cite{carlitz1987reflections} eternal thermal flux, where $\alpha\sim t^{-1/2}$, more familiar in null coordinate $v=t+z$ (as we will see in the next subsection) as $\alpha(v)=-(1/\kappa)\sqrt{-1/(\kappa v)}$. Since this does not have asymptotically constant acceleration, we postpone this to Section~\ref{sec:trscwf} and turn to $n=1$, i.e.\ $w\sim t$ hence $\alpha\sim\,$const (in agreement with Eq.~\eqref{eq:fzp1}). 

However, $\alpha=\,$const cannot be all the story, otherwise $\dot\alpha=0$ and the energy flux is zero. We must have an {\it approach\/} to constant acceleration, in a particular way, to obtain constant flux produced. From Eq.~\eqref{eq:fzp1} we see that 
\be 
\dot\alpha(t)=\ddot w\sim w^{-3}\sim t^{-3}\,, 
\ee 
for $n=1$, hence thermal flux is only produced when the acceleration asymptotically approaches constant as 
\be 
\alpha(t)=\kappa+\mathcal{O}\left(t^{-2}\right)\ . \label{eq:alt2} 
\ee 
Thus, if one wants to connect two constant acceleration asymptotic regimes, and yet obtain non-zero flux plateaus, one must ensure that Eq.~\eqref{eq:alt2} is satisfied by the asymptotes. 

Recall that Eq.~\eqref{eq:fzp1} holds for $\dot z\to 1$. What if we change the direction of motion such that $\dot z\to -1$? This flips all signs in Eq.~\eqref{eq:zwexp}, and we obtain
\be 
-12\pi F(t)\approx -\frac{\ddot w}{4w}\qquad (\dot z\to -1)\ . \label{eq:fzm1}
\ee 
The only solution is $w\sim t^n$ with $n=1$, and so $\ddot w\sim t$, implying $\alpha=\kappa+\mathcal{O}(t^2)$. This violates our assumption of asymptotically 
constant acceleration for asymptotic states $t\to\pm\infty$ 
(even for $t\to 0$ this implies the celerity 
$w\to 0$, invalidating the expansion in Eq.~(\ref{eq:zwexp})). 

Thus, a mirror with velocity asymptotically approaching the speed of light, and with acceleration approaching asymptotically constant as $1/t^2$, can produce a constant flux plateau. 

%%%%%%%%%%

\subsection{Transition in Advanced Time} 

Changing from spacetime coordinates $t$, $z$ to null coordinates $u=t-z$, $v=t+z$, the acceleration is given by 
\be 
\alpha(v)=-\frac{1}{2}\frac{f''}{(f')^{3/2}}\ , 
\ee 
where $f(v)$ is the label function for the null coordinate $u$, and $f'(v)=df/dv$. That is, $f'$ is the equivalent of the ``velocity'' of how one coordinate changes with respect to the other. The energy flux is 
\be 
-12\pi F(v)=\frac{\alpha'}{(f')^{3/2}}
=-\frac{1}{2(f')^2}\,\left[\frac{f'''}{f'}-\frac{3}{2}\left(\frac{f''}{f'}\right)^2\,\right]\ , \label{flux}
\ee 
and so again it is clear that uniformly constant acceleration will produce no energy flux. 

For $f\sim v^n$, we see that constant acceleration corresponds to $n=-1$. Seeking the behavior of the approach to the asymptote, for $f\sim v^{-1}+v^m$, we find a flux 
plateau for $m=-3$, so 
\bea 
f(v)&=&-\frac{1}{\kappa^2 v}-\frac{c}{\kappa^4v^3} \label{eq:fvv3}\ , \\ 
\alpha(v)&\approx&\kappa+\frac{3c}{2\kappa v^2} \label{eq:alv2}\ , \\ 
F(v)&\approx&\frac{\kappa^2}{48\pi}\cdot12c\ . \label{eq:fluxv2} 
\eea  
Thus, this can produce a thermal flux despite approaching constant acceleration asymptotically, as $v\to -\infty$. Since $dv/dt=1+\dot z$, then $v\to -\infty$ corresponds to $t\to -\infty$ for $\dot z\to +1$, and this behavior also holds for $v\to+\infty$, $t\to+\infty$, $\dot z\to+1$. Note that having a (non-zero) flux plateau from asymptotically constant acceleration is not possible at some finite $v$.

Finally, the correspondent distribution of created particles, the particle spectrum, is given by
\be
N_{\omega\omega'}=\left|\beta_{\omega\omega'}\right|^2\ ,
\ee
where $\beta_{\omega\omega'}$ is the appropriate Bogoliubov coefficient, with $\omega'$ and $\omega$ the frequency of the ``in'' and ``out'' modes, respectively \cite{Birrell:1982ix}. One can show that, the Bogoliubov coefficient $\beta_{\omega\omega'}$ is derived in terms of $f(v)$ as
\be
\beta_{\omega\omega'}=\frac{1}{2\pi}\sqrt{\frac{\omega'}{\omega}}\int\,dv\,e^{-i\omega'v-i\omega f(v)}\ ,
\ee
where the particular case considered determines the integration limits.

%%%%%%%%%%

\section{Transition Connecting Carlitz-Willey Accelerations}\label{sec:trscwf} 

If we already have an acceleration that by itself would give eternal thermal flux with temperature $T_1=\kappa_1/(2\pi)$, i.e.\ the Carlitz-Willey (CW) mirror \cite{carlitz1987reflections}, we can also consider a transition that connects it to a thermal flux with a different temperature $T_2=\kappa_2/(2\pi)$. This is a warm up (pun intended) for a transition connecting two horizons, in the next section, that differ in their thermal states. 

For the CW mirror  
\be 
\alpha(v)=-\frac{1}{2}\sqrt{-\frac{\kappa}{v}}\ . \label{eq:cw}
\ee 
In the time coordinate this gives asymptotic exponential acceleration, $\alpha(t\to\pm\infty)=-(\kappa/2)e^{\pm\kappa t}$, but we will remain in the advanced null coordinate $v$ since Eq.~\eqref{eq:cw} holds for all $v$. 
The label function $f(v)$ is
\be 
f(v)=-\frac{1}{\kappa}\,\ln(-\kappa v)\ , \label{eq:cwf}
\ee 
which will have a horizon at $v=0$.

Interestingly, despite the eternal thermal flux there is no past horizon. This ties back to the analysis of the previous section, in that for $\dot z\to-1$ the condition $\ddot w/w=\,$const of Eq.~\eqref{eq:fzm1} says the precise form $\alpha\sim v^{-1/2}$ gives constant flux, alternately for $\dot z\to+1$ the condition $w^3\ddot w=\,$const of Eq.~\eqref{eq:fzp1} gives the same form. 

The explicit form for $f(v)$ in Eq.~\eqref{eq:cwf} is quite interesting. It can interpreted as 
\be 
f(v) \sim \frac{1}{r}\ln\left([-\kappa v]^r\right) \sim 
\frac{1}{r}\left[\ln(-\kappa v)+\dots+\ln(-\kappa v)\right]\ , 
\ee 
where there are $r$ copies in total. As we will see, we can then displace each copy, as $\ln[-\kappa_i(v-v_i)]$, to obtain transitions between each thermal plateau.

Thus, to get a transition between two equilibrium temperatures, we can write (as one simple choice)
\be 
f(v)=a_1 \ln(-\kappa_1 v)+a_2 \ln[\kappa_2(d-v)]\ , \label{eq:trscwfgen} 
\ee 
where $d$ is some positive constant. We will find that $d$ does not affect the asymptotic temperatures, as the mirror never reaches $v=d$ due to the horizon at $v=0$.

One could instead shift both coordinates, i.e.\ use $-\kappa_1(v-d_1)$ and $\kappa_2(d_2-v)$ to end up with two horizons at $v=d_1$, $d_2<0$, with the mirror traversing between the horizons. This will be similar to the Schwarzschild-de Sitter case -- we return to this 
in the next section. Finally, we can string together further copies, e.g.\ $v$, $d_3-v$, $d_4-v$, etc.\ but this will have a limited scope, shifting the coefficients but not the asymptotic thermal states. (This generalizes significantly the setup in \cite{2102.00158}.)

Returning to Eq.~\eqref{eq:trscwfgen}, we find that 
\be 
f(v)=-\frac{1}{\kappa_2}\,\ln(-\kappa_1 v)+
\frac{\kappa_1-\kappa_2}{\kappa_1\kappa_2}\,\ln\left[\kappa_2(d-v)\right]\ . \label{CWtranstraj}\ee 
This gives an asymptotic spectrum equivalent to a CW mirror with $\kappa_1$ as $v\to -\infty$, and one with $\kappa_2$ as $v\to0$. Of course, for $\kappa_1=\kappa_2$ we simply have the standard CW case. This results in the particle spectrum, 
\be
N_{\omega\omega'}=\frac{\omega'd^2}{2\pi\kappa_2}\frac{e^{\frac{\pi\omega}{\kappa_2}}}{e^{\frac{2\pi\omega}{\kappa_2}}-1}\left|U\left(1-\frac{i\omega}{\kappa_2},2-\frac{i\omega}{\kappa_1},id\omega'\right)\right|^2,\label{U2}
\ee
which demonstrates an explicit Planck distribution indicative of the radiative horizon times the modulus of a confluent hypergeometric function. That is, at high frequencies (those seen by a distant observer) $\omega'\gg\omega$, the particle spectrum is Planckian to leading order,
\be
N_{\omega\omega'}=\frac{1}{2\pi\kappa_2\omega'}\frac{1}{e^{2\pi\omega/\kappa_2}-1}\ . \label{eq:bbody} 
\ee
We find at low frequencies, $\omega'\ll \omega$, an expected Planck spectrum,
\be
N_{\omega\omega'} = \frac{1}{2\pi\kappa_1\omega'}\frac{1}{e^{2\pi\omega/\kappa_1}-1}\ . \label{eq:bbodylow} 
\ee
As we shall see, the non-equilibrium transition between thermal states is full of new information, as highlighted by the confluent hypergeometric function in Eq.~(\ref{U2}). Likewise, spectral variations on the characteristic transition between thermal states will also be present between horizons, as illustrated in the following section.

%%%%%%%%%%

\begin{figure}[hp]
%\centering
%\captionsetup{justification=centering}
%\begin{subfigure}{0.5\textwidth}
  \includegraphics[width=0.95\linewidth]{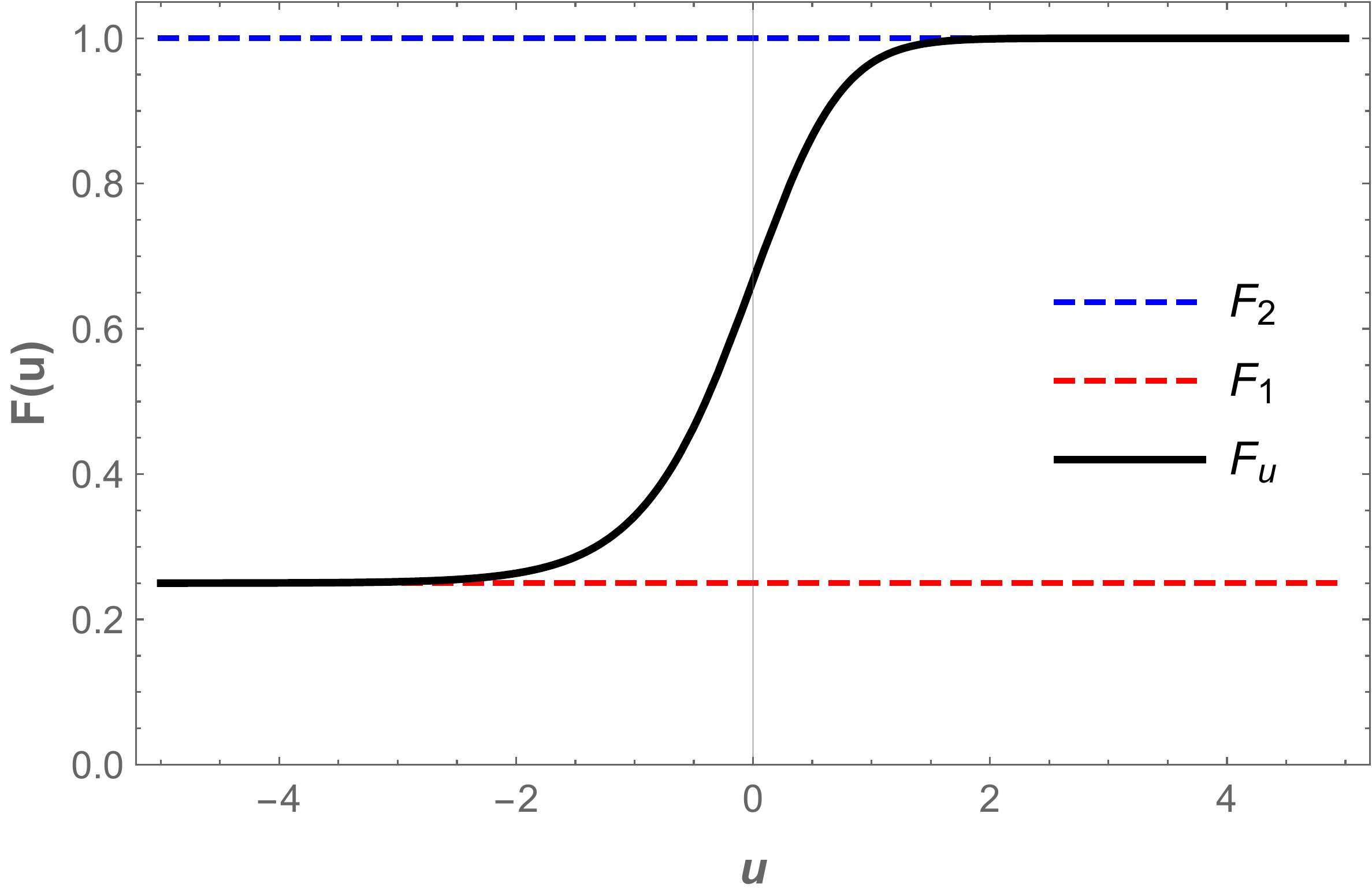}
  %\put(-170,145){a)}
  %\caption{A subfigure}
%\label{fig:4.2a}
%\end{subfigure}%
 %\begin{subfigure}{0.5\textwidth}
%\centering
%\includegraphics[width=0.8\linewidth,angle=90]{UA_spacetime.pdf}
%\put(-180,120){}
  %\caption{A subfigure}
 %\label{fig:sub2}
%\end{subfigure} 
\caption{Energy flux Eq.~(\ref{flux}) from the trajectory Eq.~(\ref{CWtranstraj}) plotted as a function of retarded time $u$, demonstrating asymptotic plateaus as measured by a distant observer, with a monotonic transition. The energy flux has been normalized by its final constant asymptotic thermal value, $F_2=\kappa_2^2/(48\pi)$. The mirror 
(here with $\kappa_1=\kappa_2/2$) ends up emitting hotter radiation in its final asymptotic state relative to its initial asymptotic state.}
\label{fig1}
\end{figure}

%%%%%%%%%%

\section{Between Two Horizons} \label{sec:twohor} 

The previous section had one horizon at $v=0$ and otherwise the mirror traversed from past infinity. Another interesting case is that of two horizons. Let us write
\be
\begin{split}
f(v)=&a_1\ln[\kappa_1(v-d_1)]+a_2\ln[\kappa_2(d_2-v)]\\
&+\sum_{j=3}^N a_j\,\ln\left[\kappa_j|v-d_j|\right]\ .\label{eq:fsum}
\end{split}
\ee
This has horizons at $v=d_1$, $d_2$ (where $d_1$ will be related to the past horizon, and $d_2$ to the future horizon) so the mirror motion is 
restricted to $v\in[d_1,d_2]$, with $d_{j>2}$ lying outside this range. 

Why include these extra factors in the summation? Our solution without them is acceptable, and we get acceleration of the thermal form as we approach each horizon: 
\bea  
\alpha(v\to d_1^+)&\to& -\frac{1}{2}\sqrt{\frac{1}{a_1(v-d_1)}}\\ 
\alpha(v\to d_2^-)&\to& -\frac{1}{2}\sqrt{\frac{1}{a_2(d_2-v)}}\ , 
\eea 
where we take $a_1=1/\kappa_1$, $a_2=1/\kappa_2$ to get the usual temperature behavior. But, if we consider the accelerated boundary correspondences with general spherically symmetric static spacetimes with metric functions given by
\be 
f(r)=1+\sum_{k=m}^nc_kr^k\ , 
\ee 
then the horizons of these spacetimes (and hence the horizons of the mirror trajectory) are determined by a degree $N=\max[n-m,n]$ polynomial. (This is easily generalized to non-integer powers by redefining variables.)

The simplest two-horizon situation is the de Sitter case,
\be
f(r)=1-\frac{\Lambda}{3}r^2\ ,
\ee
where $\Lambda$ is the cosmological constant. Then, we have one term, with $m=n=2$, and the horizons are determined by an order $N=2$ polynomial. This particular case is detailed in \cite{Good:2020byh}.

An interesting situation is the Schwarzschild-de Sitter case,
\be
f(r)=1-\frac{2M}{r}-\frac{\Lambda}{3}r^2\ ,
\ee
where $M$ is the black hole mass. Then, we have two terms, with $m=-1$ and $n=2$, and the horizons are determined by an order $N=3$ polynomial. The third solution will lead to a $j=3$ term in Eq.~\eqref{eq:fsum}. If we had $N=4$, we would have $j=3$ and $j=4$ terms in Eq.~\eqref{eq:fsum} and so on.

{\it However\/}, the trajectory is trapped within the two horizons (generally taken here to be at $d_1$ and $d_2$), and the asymptotic behaviors are independent of the other terms that cannot be accessed, but also corresponding to ``horizons''. What these terms in $f(v)$ will do is affect the details of how the non-thermal transition between the asymptotic thermal states occurs, and hence the non-asymptotic spectrum. This apparently suggests that the characteristics of the radiation can probe ``superhorizon'' conditions (beyond the least negative $v$ horizon or beyond the most negative $v$ horizon). In the Schwarzschild-de Sitter case, the future horizon corresponds to the (Schwarzschild) black hole horizon, and the past horizon corresponds to the (de Sitter) cosmological horizon, meaning effects apparently come from 
inside the Schwarzschild horizon and outside the de Sitter 
horizon). Really though, the form of $f(v)$ within the horizons is being determined by the unrealized metric polynomial solutions that give rise to the inaccessible ``ghost horizons''.

The particle spectrum in transition between the two thermal states will be described by a confluent hypergeometric function of $N-1$ variables, since $e^{i\omega f(v)}$ simply factorizes into $N-1$ terms of the form $(v-d_j)^{i\omega a_j}$. In the Schwarzschild-de Sitter case, the extra term in Eq.~\eqref{eq:fsum} imposes a confluent hypergeometric function of two variables, $\Phi_1$, also known as the Humbert function or the Humbert series, in the Bogoliubov coefficient. One of the variables involves the distance between the horizons and the mode frequency, as if there were no third term in $f(v)$. The other variable also involves the distance between the horizons, and $d_3$ from the mixing term.

Finally, we can extend the solution of accelerated boundary correspondence with Schwarzschild-de Sitter spacetime as given in \cite{2109.04147} (which approximates by neglecting the third term) to include the full solution with all three terms present. The approximated Bogoliubov coefficient considering only two terms involves the confluent hypergeometric function of one variable. The asymptotic behaviour of ${}_1F_1(a,b;x)$ as $x\to\infty$ is
\be 
{}_1F_1(a,b;x)\approx\left(\frac{x^{a-b}e^x}{\Gamma(a)}-\frac{(-x)^{-a}}{\Gamma(b-a)}\right)\,\Gamma(b)\ ,
\ee 
which corresponds to the high frequency limit, as seen by an inertial observer at infinity.

Let us now write the function that parameterizes the analog Schwarzschild-de Sitter mirror trajectory as \cite{2109.04147}
\be
\begin{split}
f(v)=&-\frac{1}{\kappa_B}\ln\left[\Bar{\kappa}_B(v_B-v)\right]+\frac{1}{\kappa_C}\ln\left[\Bar{\kappa}_C(v-v_C)\right]\\
&-\frac{1}{\kappa_0}\ln\left[\Bar{\kappa}_0(v+v_0)\right]\ ,
\end{split}
\ee
where $v_0=v_B+v_C$ and $\kappa_0^{-1}=\kappa_C^{-1}-\kappa_B^{-1}$. All the parameters of the trajectory have a one-to-one correspondence with the parameters of the Schwarzschild-de Sitter spacetime. As far as the mirror is concern, $v_i\ (i=B,C)$ describe the acceleration horizons of the trajectory, $\kappa_i\ (i=B,C)$ describe the acceleration parameters of the trajecory, and $\Bar{\kappa}_i\ (i=B,C,0)$ are, to this extent, irrelevant constants. See \cite{2109.04147} for details. A Penrose diagram of Minkowski spacetime with superposition of a mirror trajectory of this type can also be found there.
The exact Bogoliubov coefficient considering the three terms involves the confluent hypergeometric function of two variables. The explicit expression is
\be
\begin{split}
\beta_{\omega\omega'}=&\frac{1}{2\pi}\sqrt{\frac{\omega'}{\omega}}\,\frac{\Bar{\kappa}_B^b\Bar{\kappa}_C^{-c}}{\Bar{\kappa}_0^{b-c}\kappa'}\,\delta^{b-c}e^{-i\omega'v_B}\\
&\cdot\frac{\Gamma(1+b)\Gamma(1-c)}{\Gamma(2+b-c)}\Phi_1(1+b,b-c,2+b-c;\delta,a)\ , \label{beta}
\end{split}
\ee
where we have defined
\be
a\equiv\frac{i\omega'}{\kappa'}\quad,\quad
b\equiv\frac{i\omega}{\kappa_B}\quad,\quad
c\equiv\frac{i\omega}{\kappa_C}\quad,\quad
\ee
with $\kappa'=(v_B-v_C)^{-1}$, and\footnote{One can see that in the limit $\delta\to0$ (with $v_B\neq v_C$), $\Bar{\kappa}_0\to0$, and $$\frac{\delta}{\Bar{\kappa}_0}\to\frac{1}{\kappa'}\ .$$ Taking this into account one can recover Eq.~(32) in \cite{2109.04147} from our Eq.~(\ref{beta}).} 
\be
\delta\equiv\frac{1}{(v_B+v_0)\kappa'}=\frac{v_B-v_C}{2v_B+v_C}\ .\label{eq:d}
\ee
The asymptotic behaviour of $\Phi_1(a,p,b;q,x)$ as $x\to\infty$ is
\be 
\Phi_1(a,p,b;q,x)\approx\left(\frac{(1-q)^{-p}x^{a-b}e^x}{\Gamma(a)}-\frac{(-x)^{-a}}{\Gamma(b-a)}\right)\,\Gamma(b)\ , 
\ee
which again corresponds to the high frequency limit, as seen by an inertial observer at infinity. Note that in this limit the asymptotic behaviour of ${}_1F_1(a,b;x)$ and $\Phi_1(a,p,b;q,x)$ look very similar. As expected, both expressions match when the second variable vanishes, i.e.\ $q=0$ implies no third term in $f(v)$.

The particle spectrum splits in the $\omega'\gg\omega$ regime as
\be
N_{\omega\omega'}=N_B+N_C+N_{BC}\ ,\label{SdSb2}
\ee
where 
\be
N_B=\frac{1}{2\pi\kappa_B\omega'}\frac{1}{e^{2\pi\omega/\kappa_B}-1}\ ,
\ee
\be
N_C=\frac{1}{2\pi\kappa_C\omega'}\frac{1}{e^{2\pi\omega/\kappa_C}-1}\ ,
\ee
and
\be
\begin{split}
N_{BC}=&-\frac{1}{4\pi^2\omega\omega'}(1-\delta)^{-\frac{i\omega}{\kappa_0}}\left(\frac{\omega'}{\kappa'}\right)^{-\frac{i\omega}{\kappa}}e^{-\frac{\pi\omega}{2\kappa}-\frac{i\omega'}{\kappa'}}\\
&\cdot\Gamma\left(1+\frac{i\omega}{\kappa_B}\right)\Gamma\left(1+\frac{i\omega}{\kappa_C}\right)+\mathrm{c.c.}\ ,
\end{split}
\ee
with $\kappa^{-1}=\kappa_B^{-1}+\kappa_C^{-1}$. This term indicates the particle creation by the accelerated boundary correspondence with Schwarzschild-de Sitter spacetime at intermediate times. It demonstrates a non-thermal particle distribution between the early-time and late-time thermal equilibria. It shows the effect of the acceleration horizons and the interaction behaviour between them resulting in the non-thermal spectrum. The particle production is rather non-trivial at the intermediate times of the trajectory, and this constitutes a new aspect in the radiation results when several physical horizons are present, demonstrating two Planck distribution `intercommunication' to end up with an overall out-of-equilibrium spectrum. In this case, this complicated cross term indicates the exact form of this interplay between the analog black hole and cosmological horizons of Schwarzschild-de Sitter spacetime. We can recast the cross term $N_{BC}$ in the following form using of the properties of the Gamma function $\Gamma(x)$, which reveals an interesting mixing modulated by an oscillation:
\be
\begin{split} 
&N_{BC}=-2\sqrt{N_BN_C}\\
&\cdot\cos\left[\frac{\omega'}{\kappa'}+\frac{\omega}{\kappa}\ln\left(\frac{\omega'}{\kappa'}\right)+\frac{\omega}{\kappa_0}\ln\left(1-\delta\right)-(\theta_B+\theta_C)\right]\ ,
\end{split}
\ee
where we have defined
\be
\theta_i\equiv-\frac{\omega}{\kappa_i}\gamma_E+\sum_{n=1}^{\infty}\left[\frac{\omega}{n\kappa_i}-\tan^{-1}\left(\frac{\omega}{n\kappa_i}\right)\right]\ (i=B,C)\ ,
\ee
with $\gamma_E$ the Euler-Mascheroni constant. The infinite series in $\theta_i$ as it stands is not convergent. As an approximation, we can expand the function $\tan^{-1}(x)$ to see that, at leading order,
\be
\sum_{n=1}^{\infty}\left[\frac{\omega}{n\kappa_i}-\tan^{-1}\left(\frac{\omega}{n\kappa_i}\right)\right]\sim\mathcal{O}(\omega^3)\ ,
\ee
so, in the $\omega'\gg\omega$ regime (with $\omega\to0$), we can neglect this term and write
\be
\theta_i\approx-\frac{\omega}{\kappa_i}\gamma_E\ .
\ee 

\begin{figure}[ht]
    \centering
    \includegraphics[width=1.00\linewidth]{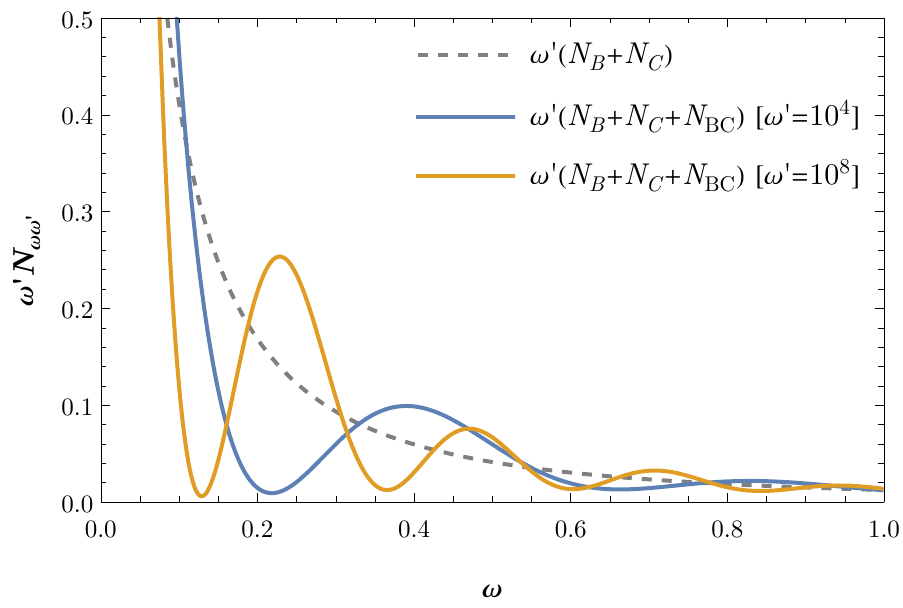}
    \caption{The particle spectrum $N_{\omega\omega'}$ for the analog Schwarzschild-de Sitter mirror, normalized by $\omega'$, is plotted as a function of $\omega$. The mixing term $N_{BC}$ between the two horizons causes an interference pattern (solid curves, for two different values of $\omega'$) modulating the unmixed Planck spectrum of $N_B+N_C$ (dashed curve). In this plot, we have taken the following numerical values for the trajectory: $v_B\simeq-0.20$, $v_C\simeq-1.89$, $\kappa_B\simeq4.80$, $\kappa_C\simeq0.89$.}
    \label{fig:sdsmix}
\end{figure}

The results are shown in Fig.~\ref{fig:sdsmix}, showing that the mixing term has a significant effect on the particle spectrum, with an explicit oscillation modulating the separate Planck spectra sum. Since $N_B$, $N_C$, and $N_{BC}$ all have prefactors of $1/\omega'$, we plot $\omega'(N_B+N_C)$ and $\omega'(N_B+N_C+N_{BC})$ to present results for two different values of $\omega'$ on the same scale. Note that, as expected, higher $\omega'$ increases the oscillation frequency. Thus the ``interference'' between the horizons yields a physically significant effect.

%%%%%%%%%%

%%%%%%%%%%

As expected, we see that when $\kappa_C=0$ (which is the analog of the Schwarzschild limit), the particle spectrum is the usual thermal spectrum, $N_B$, since $N_C$ and $N_{BC}$ vanish.

\section{Transition Connecting Horizons with Davies-Fulling Acceleration} \label{sec:df} 

As an another example connecting horizons, let us now consider the unusual transition between the two horizons of the `null-self-dual' or `Davies-Fulling' trajectory,
\be
f(v) = -\frac{1}{\kappa}\ln (2-e^{\kappa v})\ ,
\ee
which is the light-speed case \cite{1711.09922} of the drifting counterpart in \cite{Good:2017kjr}, first suggested by Davies-Fulling \cite{Davies:1977yv} and then investigated at late-times \cite{Birrell:1982ix}. Solving for the overall particle spectrum we find an interesting form,
\be 
N_{\omega\omega'} = \frac{1}{2\pi \kappa (\omega-\omega')} \left(\frac{1}{e^{2 \pi \omega'/\kappa}-1}-\frac{1}{e^{2 \pi  \omega/\kappa}-1}\right).\label{DFb2}
\ee
This is a sum of Planck factors with no apparent mixing between the horizons, presumably due to the self-dual nature. That is, the characteristic dynamical trait of the Davies-Fulling null-self-dual trajectory is time-reversal symmetry on the trajectory motion, $z(t) = z(-t)$. The exponential proper acceleration in space (as the independent variable) or inverse linear dependence on proper time results in a side-symmetric spectrum and dual energy flux which is the same on both sides of the mirror, hence the reasonably apt name, `self-dual'. Eq.~(\ref{DFb2}) is correct even in the null limit of the drifting case in \cite{Good:2017kjr}. Compare this no-mix behavior to the CW-CW transition spectrum in Sec.~\ref{sec:trscwf} which also has an explicit Planck spectra, but includes a mixing effect due to the confluent hypergeometric function modulus squared.

In the high frequency limit, $\omega'\gg \omega$, the radiation reduces to the expected and explicit Planck distribution, i.e.\ Eq.~(\ref{eq:bbody}), in full agreement with Davies-Fulling late-time results. Symmetrically, in what we may call the `low frequency' limit, $\omega'\ll \omega$, the radiation simplifies to
\be 
N_{\omega\omega'} = \frac{1}{2 \pi \kappa \omega} \frac{1}{e^{2 \pi \omega' / \kappa} -1}\ , \label{PLANCK2}
\ee
demonstrating a temperature $T = \kappa/(2\pi)$ as well. Notice the difference in prime location compared with Eq.~(\ref{eq:bbodylow}).

The energy flux straightforwardly gives a plateau at late retarded times using $p(u)=(1/\kappa)\ln(2-e^{-\kappa u})$, which corroborates the high frequency result.  However, at early times, no such plateau exists. Where then is the thermal flux indicative of the Planck spectrum in Eq.~(\ref{PLANCK2})?  The early-time missing plateau exists, but on the \textit{left} side of the mirror, as can be found by symmetry, replacing the form of $f(v)\leftrightarrow p(u)$ and investigating early retarded time emission in $-24\pi F(u)=p'''/p'-(3/2)(p''/p')^2$.

%%%%%%%%%%

\section{Conclusions} \label{sec:concl} 

The non-equilibrium evolution between horizons and thermal states is a rich subject, associated with non-intuitive but fundamental dynamics and radiative symmetries with fascinating consequences, like curvature correspondences \cite{DeWitt:1975ys}, negative-energy particles \cite{Ford:1997hb} or information loss \cite{Fabbri,Chen:2015bcg,Akal:2020twv,Good:2021asq,Sato:2021ftf}. One benefit to deriving the radiation emitted by thermal horizons is the consistency check afforded by the appearance of a Planck spectrum in the appropriate frequency regime, absent in other contexts like the inertial horizon \cite{Good:2020uff}, eternal constant acceleration \cite{Good:2021iny}, or extremal horizons \cite{good2020extreme}. 

In the investigation of such asymptotic equilibria we have found four novel solutions corresponding to transitions between horizons from time-dependent accelerated motions:  
\begin{itemize} 
    \item CW-CW transition spectrum; Eq.~(\ref{U2}); Sec.~\ref{sec:trscwf}.
    \item S-dS transition spectrum; Eq.~(\ref{SdSb2}); Sec.~\ref{sec:twohor}. 
    \item D-F transition spectrum; Eq.~(\ref{DFb2}); Sec.~\ref{sec:df}. 
    \item CW-CA transition spectrum; Eq.~(\ref{CWCAb2});  App.~\ref{sec:trsconcw}.
    \end{itemize} 
(In addition we have studied asymptotic thermal flux and particle spectrum from transitions between constant acceleration horizons, Sec.~\ref{sec:trscon}, Eq.~\eqref{eq:fvv3}.) 
    
There are important physical insights gleaned from each of the major solutions: The CW-CW spectrum in Sec.~\ref{sec:trscwf} unravels the fact of having physical evolution from an asymptotic thermal flux with no horizon, to a separate asymptotic thermal flux with a higher or lower temperature. This not only demonstrates the necessarily detached relation between the existence of a horizon and thermality, but the freedom to step-up (or down) in temperature within a trajectory defined globally. (And has some interesting asides on ``superhorizon'' structure.) 

The most important advancement in Sec.~\ref{sec:twohor}, and a distinctive highlight of this paper, is the solution for accelerated boundary correspondence with Schwarzschild-de Sitter spacetime.  We have been able to analytically derive the radiation emitted by the analog black hole and cosmological horizons, which is a particularly challenging transition particle spectrum that was previously considered intractable \cite{2109.04147}. The results are in agreement with what was qualitatively expected, but the exact and precise result quantitatively incorporates an oscillation that characterizes the mixing of asymptotic thermal states. An important and interesting aspect to address in future investigations concerns the detectability of this horizon ``interference'' by an observer. A first approach may be to reanalyze how Unruh-DeWitt detectors respond in these scenarios where different physical horizons are present. It is also worth mentioning that alternative configurations of the dynamical Casimir effect exhibit an interference phenomenon \cite{Silva:2015yta}, which might be more accessible experimentally. However, in these cases the cause is different from having an accelerated boundary with two horizons.

A direct comparison to Sec.~\ref{sec:df} leads to the remarkable understanding of the salient physical simplicity of essentially no mixing from the transition if the acceleration is that of the Davies-Fulling trajectory. There the particle spectrum from the transition between horizons is just a straight sum of Planck factors for each of the horizons. What's more, the hidden plateau on the left-side of the mirror has implications in 1+3 dimensions \cite{Candelas:1977zza,Frolov_1979,Frolov:1980pg}, where both sides of the mirror are needed for generalization \cite{Good:2021ffo,Zhakenuly:2021pfm}. 

An interesting result is treated in Appendix~\ref{sec:trsconcw}. This is the analytic spectral tractability of a transition connecting constant acceleration to CW that in the end reveals the Planck factor associated with the thermal plateau in energy flux. Perhaps the most interesting physical implication is not that energy flux can be independent of branch cuts that result in multiple horizons within a trajectory defined globally, but that there exist negative energy flux thermal plateaus (e.g.~\cite{Davies:1982cn,Walker_1982,Walker:1984ya,walker1985particle,Ford:2004ba}) characterized by Planck factors.

Already established gravitational analogue models like the accelerated moving mirror, as well as the new ones, continue to deliver theoretical \cite{Lynch:2015lra,Akal:2022qei} and experimental insights \cite{Lynch:2019hmk,AnaBHEL:2022sri} into the nature of acceleration radiation, horizons, and thermality. Understanding how higher dimensions may play a role with respect to transitions, and providing an in-depth analysis of the physical implications in curved spacetimes for the trajectories studied would be an interesting subject for investigation.

In the context of Schwarzschild-de Sitter spacetime specifically, further investigation is needed to recover and reinterpret the results of the analog mirror but now in the curved spacetime system. In the case of Schwarzschild spacetime this was achieved in \cite{Good:2016oey}, but in the Schwarzschild-de Sitter case this is more complicated. Note that the accelerated boundary system is asymptotically flat, as is the Schwarzschild spacetime, but the Schwarzschild-de Sitter spacetime is not. There the correspondence with the curved spacetime is not as straightforward. We defer the details of such an investigation to a future work.

%%%%%%%%%%

\acknowledgments 

We thank Byron \cite{Byron} for insight into creation/annihilation motivating the title. DFS acknowledges support from the Mathematical Physics group at the University of Burgos. This work has been partially supported by Agencia Estatal de Investigación (Spain) under grant PID2019-106802GB-I00/AEI/10.13039/501100011033, by the Regional Government of Castilla y León (Junta de Castilla y León, Spain), and by the Spanish Ministry of Science and Innovation MICIN and the European Union NextGenerationEU/PRTR. Funding comes in part from the FY2021-SGP-1-STMM Faculty Development Competitive Research Grant No.\ 021220FD3951 at Nazarbayev University, the Energetic Cosmos Laboratory, and the U.S. Department of Energy, Office of Science, Office of High Energy Physics, under contract no.\  DE-AC02-05CH11231.

%%%%%%%%%%

\appendix 

\section{Transition Connecting Constant and Carlitz-Willey Accelerations} \label{sec:trsconcw} 

For comparison and completeness, let us consider a transition that asymptotically approaches constant acceleration in the past in the manner needed to obtain a energy flux plateau, and asymptotically approaches the CW thermal behavior in the future. This can be realized with the mirror trajectory  
\be 
f(v)=\frac{1}{\kappa}\sinh^{-1}\left(-\frac{1}{\kappa v}\right)\ , \label{CWtoConstant1} 
\ee 
where there are two branches, either the left-hand side, $v\in[-\infty,0]$ or the right-hand side, $v\in[0,\infty]$. 

\begin{figure*}[ht!]
    \centering
    \begin{subfigure}[t]{0.4\textwidth}
        \centering 
        \includegraphics[height=2in]{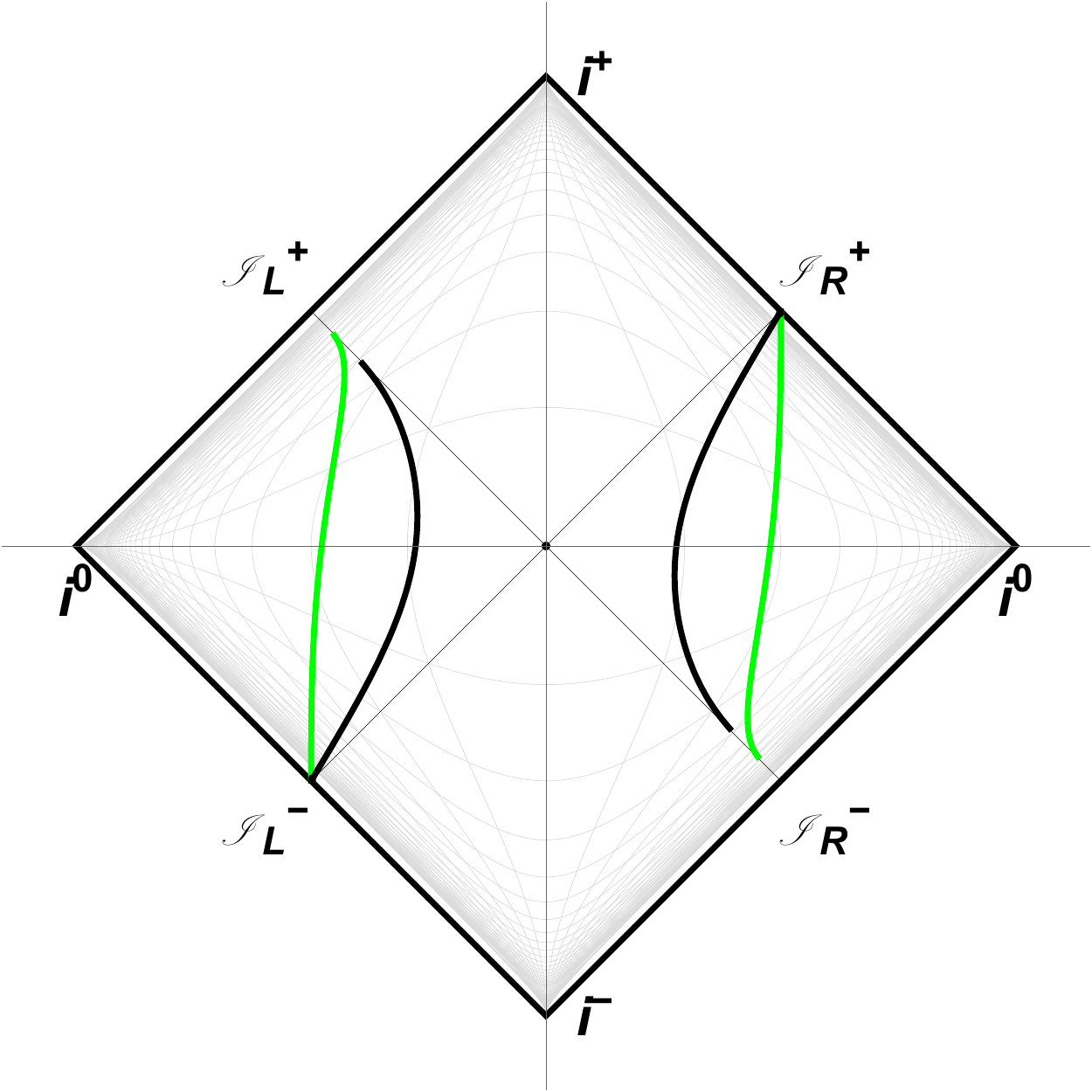} 
        \caption{Trajectories of the mirror of Eq.~(\ref{CWtoConstant1}), for $\kappa =1,2$ (green, black).} 
   \end{subfigure} 
    ~ 
    \begin{subfigure}[t]{0.5\textwidth}
        \centering 
        \includegraphics[height=2in]{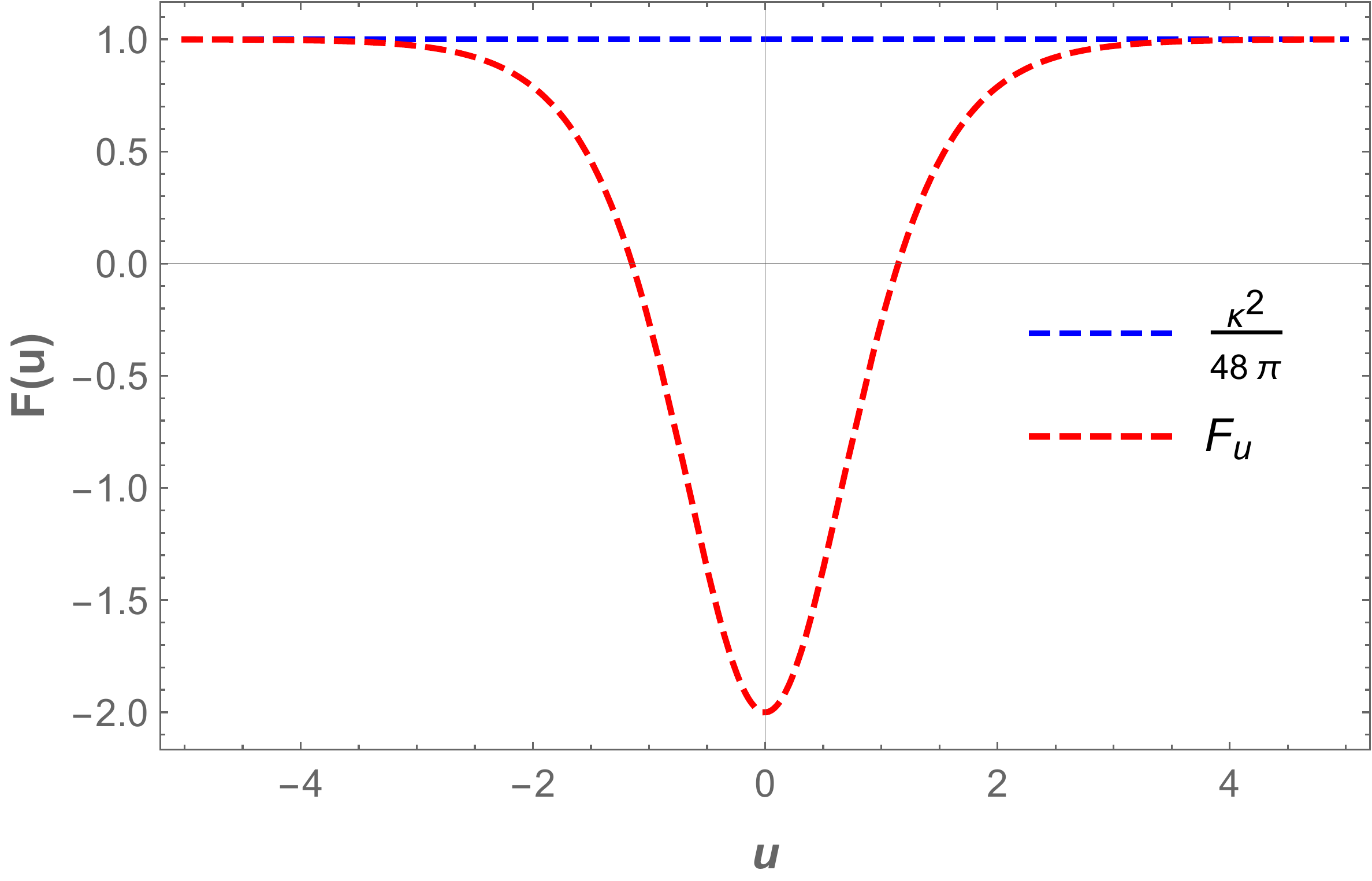} 
        \caption{Energy flux from the mirror of Eq.~(\ref{CWtoConstant1}).} 
\label{fig:concw2b} 
    \end{subfigure} 
\caption{Spacetime diagram [left panel] and energy flux [right panel] for the mirror trajectory of Eq.~(\ref{CWtoConstant1}). The left branch on the spacetime diagram ranges from $u\in[0,+\infty]$ (from constant acceleration to CW, confusingly appearing on the right half of the energy flux plot) while the right branch on the spacetime diagram ranges from $u\in[-\infty,0]$ (from CW to constant acceleration). The energy flux (plotted as a function of retarded time $u$ to bring out the plateaus at $|\kappa u|\gg1$, i.e.\ $|\kappa v|\ll1$) exhibits both future and past asymptotic approach to a plateau as measured by a distant observer. The energy flux has been normalized by its final constant asymptotic thermal value $F_{\rm Th} = \kappa^2/(48\pi)$, and includes both branches on the plot.}\label{fig:concw2}
\end{figure*}

Regardless of branch, the energy flux $F(v)$ from this mirror, using Eq.~(\ref{flux}), is
\be 
F(v) = \frac{\kappa^2}{48\pi}\,\frac{1-2 \kappa ^2 v^2}{1+ \kappa ^2 v^2}\ . \label{flux1} 
\ee 
The energy flux is asymptotically thermal in both the past and future: $F(v)\to[\kappa^2/(48\pi)]\cdot(-2,+1)$. The unusual $-2$ factor arises because of the particular transition we chose between constant acceleration and CW. Recall that Eq.~\eqref{eq:fvv3} gives an energy flux approaching thermal with $F(v)\to[\kappa^2/(48\pi)]\cdot12c$, and the expansion of $\sinh^{-1}(x)$ in Eq.~\eqref{CWtoConstant1} gives $c=-1/6$. The two branches and energy fluxes are shown in Fig.~\ref{fig:concw2}. 

\begin{figure*}[ht!] 
    \begin{subfigure}[t]{0.48\textwidth}
        \centering 
        \includegraphics[height=2in]{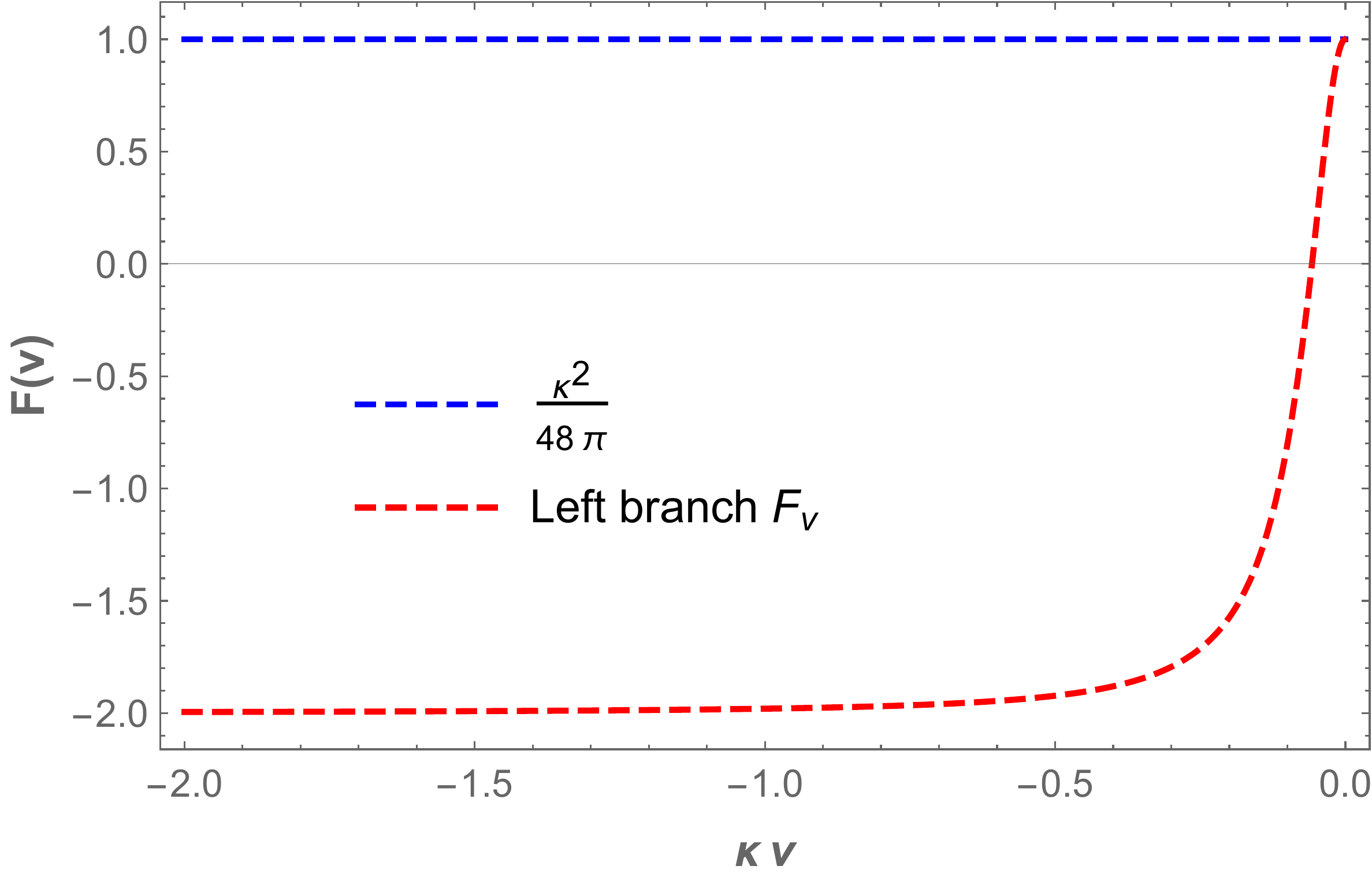} 
\caption{Left branch energy flux of the mirror of Eq.~(\ref{CWtoConstant1}), vs $\kappa v$.} 
    \end{subfigure} 
    ~ 
    \begin{subfigure}[t]{0.48\textwidth}
        \centering 
        \includegraphics[height=2in]{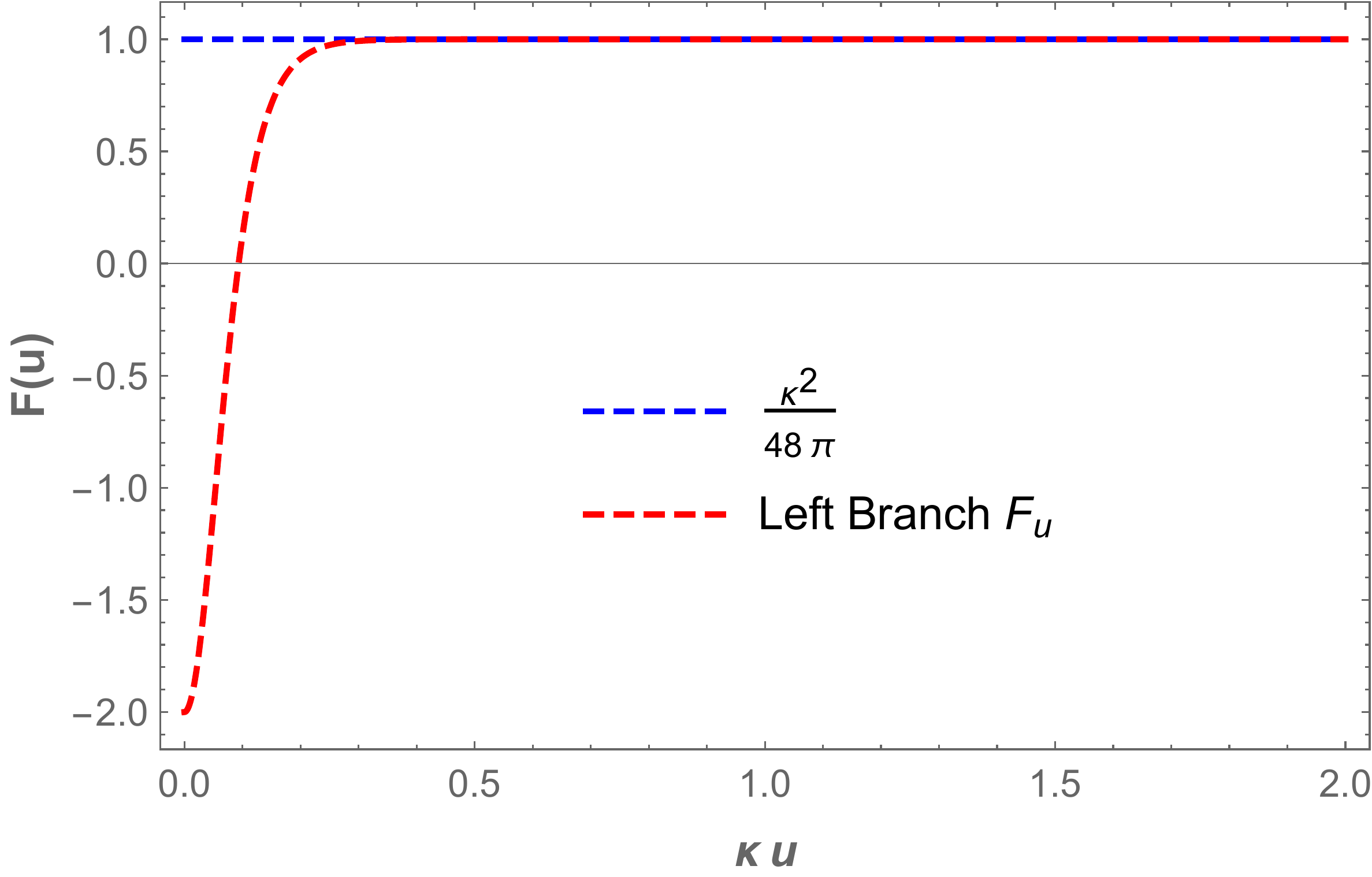} 
\caption{Left branch energy flux of the mirror of Eq.~(\ref{CWtoConstant1}), vs $\kappa u$.}
    \end{subfigure}
\caption{Energy flux is plotted vs both $\kappa v$ [left panel] and $\kappa u$ [right panel] time variables. The plots are of the same curve, simply plotted in different time variables 
to bring out the plateaus clearly. The tail (at $F/F_{\rm Th}=-2$) to large $|\kappa v|$ in the left panel {\it is\/} the short dip at $\kappa u\approx 0$ in the right panel, and the tail 
(at $F/F_{\rm Th}=+1$) to large $\kappa u$ in the right panel {\it is\/} the short 
hill at $\kappa v\approx 0$ in the left panel (and both are the left branch, i.e.\ right half, in Fig.~\ref{fig:concw2b}, shown there vs $u$ rather than $\kappa u$).}
\label{fig:concwuv}
\end{figure*}

Fig.~\ref{fig:concwuv} demonstrates each of the two regions of asymptotic thermality possessed individually by each branch (here shown for the left-hand side branch), shown in two different coordinates 
since the plateau near $u=0$ ($v\to\pm\infty$) is hard to see when plotted in $u$ (just as the one near $v=0$ ($u\to\pm\infty$) is hard to see when plotted in $v$). Indeed the energy flux is constant for $|\kappa v|\gg1$ and $|\kappa u|\gg1$ 
($v\to 0$). 

%%%%%%%%%%

%%%%%%%%%%

We can further check thermality by considering the particle spectrum. This can be calculated analytically, giving 
\be 
N_{\omega\omega'} = \frac{\omega}{16\pi^4 \kappa^2 \omega'}  \left|G_{2,4}^{4,1}\left(C\; \Big|
\begin{array}{c}
 A\\
 B\\
\end{array}
\right)\right|^2\ , \label{CWCAb2}
\ee
where $A = (1-i\omega/(2\kappa), 1+i\omega/(2\kappa))$, $B= (0,1/2, 1/2 ,1)$, and $C = -\omega'^2/(4\kappa^2)$. Here $G$ is the Meijer G-function, a general function which reduces to well-known simpler special functions as particular cases. It characterizes the particle spectrum of an inertial horizon \cite{Good:2020uff}. See also the effective action of moving mirrors freely-falling onto a black hole \cite{Sorge:2018zfd,Sorge:2019ecb}. 

In the high frequency approximation $\omega'\gg\omega$, one verifies to leading order that
\be 
N_{\omega\omega'} = \frac{1}{2 \pi \kappa \omega'}\frac{1}{e^{2 \pi \omega / \kappa} -1}\ . \label{PLANCK}
\ee 
This is precisely the thermal spectrum. 

\clearpage\bibliography{main}

\begin{thebibliography}{50}%
\makeatletter
\providecommand \@ifxundefined [1]{%
 \@ifx{#1\undefined}
}%
\providecommand \@ifnum [1]{%
 \ifnum #1\expandafter \@firstoftwo
 \else \expandafter \@secondoftwo
 \fi
}%
\providecommand \@ifx [1]{%
 \ifx #1\expandafter \@firstoftwo
 \else \expandafter \@secondoftwo
 \fi
}%
\providecommand \natexlab [1]{#1}%
\providecommand \enquote  [1]{``#1''}%
\providecommand \bibnamefont  [1]{#1}%
\providecommand \bibfnamefont [1]{#1}%
\providecommand \citenamefont [1]{#1}%
\providecommand \href@noop [0]{\@secondoftwo}%
\providecommand \href [0]{\begingroup \@sanitize@url \@href}%
\providecommand \@href[1]{\@@startlink{#1}\@@href}%
\providecommand \@@href[1]{\endgroup#1\@@endlink}%
\providecommand \@sanitize@url [0]{\catcode `\\12\catcode `\$12\catcode
  `\&12\catcode `\#12\catcode `\^12\catcode `\_12\catcode `\%12\relax}%
\providecommand \@@startlink[1]{}%
\providecommand \@@endlink[0]{}%
\providecommand \url  [0]{\begingroup\@sanitize@url \@url }%
\providecommand \@url [1]{\endgroup\@href {#1}{\urlprefix }}%
\providecommand \urlprefix  [0]{URL }%
\providecommand \Eprint [0]{\href }%
\providecommand \doibase [0]{http://dx.doi.org/}%
\providecommand \selectlanguage [0]{\@gobble}%
\providecommand \bibinfo  [0]{\@secondoftwo}%
\providecommand \bibfield  [0]{\@secondoftwo}%
\providecommand \translation [1]{[#1]}%
\providecommand \BibitemOpen [0]{}%
\providecommand \bibitemStop [0]{}%
\providecommand \bibitemNoStop [0]{.\EOS\space}%
\providecommand \EOS [0]{\spacefactor3000\relax}%
\providecommand \BibitemShut  [1]{\csname bibitem#1\endcsname}%
\let\auto@bib@innerbib\@empty
%</preamble>
\bibitem [{\citenamefont {Hawking}(1975)}]{Hawking:1974sw}%
  \BibitemOpen
  \bibfield  {author} {\bibinfo {author} {\bibfnamefont {S.}~\bibnamefont
  {Hawking}},\ }\href {\doibase 10.1007/BF02345020} {\bibfield  {journal}
  {\bibinfo  {journal} {Commun. Math. Phys.}\ }\textbf {\bibinfo {volume}
  {43}},\ \bibinfo {pages} {199} (\bibinfo {year} {1975})}\BibitemShut
  {NoStop}%
\bibitem [{\citenamefont {Fulling}(1973)}]{Fulling:1972md}%
  \BibitemOpen
  \bibfield  {author} {\bibinfo {author} {\bibfnamefont {S.~A.}\ \bibnamefont
  {Fulling}},\ }\href {\doibase 10.1103/PhysRevD.7.2850} {\bibfield  {journal}
  {\bibinfo  {journal} {Phys. Rev. D}\ }\textbf {\bibinfo {volume} {7}},\
  \bibinfo {pages} {2850} (\bibinfo {year} {1973})}\BibitemShut {NoStop}%
\bibitem [{\citenamefont {Davies}(1975)}]{Davies:1974th}%
  \BibitemOpen
  \bibfield  {author} {\bibinfo {author} {\bibfnamefont {P.~C.~W.}\
  \bibnamefont {Davies}},\ }\href {\doibase 10.1088/0305-4470/8/4/022}
  {\bibfield  {journal} {\bibinfo  {journal} {J. Phys. A}\ }\textbf {\bibinfo
  {volume} {8}},\ \bibinfo {pages} {609} (\bibinfo {year} {1975})}\BibitemShut
  {NoStop}%
\bibitem [{\citenamefont {Unruh}(1976)}]{unruh76}%
  \BibitemOpen
  \bibfield  {author} {\bibinfo {author} {\bibfnamefont {W.~G.}\ \bibnamefont
  {Unruh}},\ }\href {\doibase 10.1103/PhysRevD.14.870} {\bibfield  {journal}
  {\bibinfo  {journal} {Phys. Rev. D}\ }\textbf {\bibinfo {volume} {14}},\
  \bibinfo {pages} {870} (\bibinfo {year} {1976})}\BibitemShut {NoStop}%
\bibitem [{\citenamefont {Carlitz}\ and\ \citenamefont
  {Willey}(1987{\natexlab{a}})}]{carlitz1987reflections}%
  \BibitemOpen
  \bibfield  {author} {\bibinfo {author} {\bibfnamefont {R.~D.}\ \bibnamefont
  {Carlitz}}\ and\ \bibinfo {author} {\bibfnamefont {R.~S.}\ \bibnamefont
  {Willey}},\ }\href {\doibase 10.1103/PhysRevD.36.2327} {\bibfield  {journal}
  {\bibinfo  {journal} {Phys. Rev. D}\ }\textbf {\bibinfo {volume} {36}},\
  \bibinfo {pages} {2327} (\bibinfo {year} {1987}{\natexlab{a}})}\BibitemShut
  {NoStop}%
\bibitem [{\citenamefont {Fulling}\ and\ \citenamefont
  {Davies}(1976)}]{Davies:1976hi}%
  \BibitemOpen
  \bibfield  {author} {\bibinfo {author} {\bibfnamefont {S.~A.}\ \bibnamefont
  {Fulling}}\ and\ \bibinfo {author} {\bibfnamefont {P.~C.~W.}\ \bibnamefont
  {Davies}},\ }\href
  {https://royalsocietypublishing.org/doi/abs/10.1098/rspa.1976.0045}
  {\bibfield  {journal} {\bibinfo  {journal} {Proc. R. Soc. Lond. A}\ }\textbf
  {\bibinfo {volume} {348}},\ \bibinfo {pages} {393} (\bibinfo {year}
  {1976})}\BibitemShut {NoStop}%
\bibitem [{\citenamefont {Davies}\ and\ \citenamefont
  {Fulling}(1977)}]{Davies:1977yv}%
  \BibitemOpen
  \bibfield  {author} {\bibinfo {author} {\bibfnamefont {P.~C.~W.}\
  \bibnamefont {Davies}}\ and\ \bibinfo {author} {\bibfnamefont {S.~A.}\
  \bibnamefont {Fulling}},\ }\href {\doibase 10.1098/rspa.1977.0130} {\bibfield
   {journal} {\bibinfo  {journal} {Proc. R. Soc. Lond. A}\ }\textbf {\bibinfo
  {volume} {A356}},\ \bibinfo {pages} {237} (\bibinfo {year}
  {1977})}\BibitemShut {NoStop}%
\bibitem [{\citenamefont {Birrell}\ and\ \citenamefont
  {Davies}(1984)}]{Birrell:1982ix}%
  \BibitemOpen
  \bibfield  {author} {\bibinfo {author} {\bibfnamefont {N.~D.}\ \bibnamefont
  {Birrell}}\ and\ \bibinfo {author} {\bibfnamefont {P.~C.~W.}\ \bibnamefont
  {Davies}},\ }\href {\doibase 10.1017/CBO9780511622632} {\emph {\bibinfo
  {title} {{Quantum Fields in Curved Space}}}},\ Cambridge Monographs on
  Mathematical Physics\ (\bibinfo  {publisher} {Cambridge Univ. Press},\
  \bibinfo {address} {Cambridge, UK},\ \bibinfo {year} {1984})\BibitemShut
  {NoStop}%
\bibitem [{\citenamefont {Good}\ \emph {et~al.}(2016)\citenamefont {Good},
  \citenamefont {Anderson},\ and\ \citenamefont {Evans}}]{Good:2016oey}%
  \BibitemOpen
  \bibfield  {author} {\bibinfo {author} {\bibfnamefont {M.~R.~R.}\
  \bibnamefont {Good}}, \bibinfo {author} {\bibfnamefont {P.~R.}\ \bibnamefont
  {Anderson}}, \ and\ \bibinfo {author} {\bibfnamefont {C.~R.}\ \bibnamefont
  {Evans}},\ }\href {\doibase 10.1103/PhysRevD.94.065010} {\bibfield  {journal}
  {\bibinfo  {journal} {Phys. Rev. D}\ }\textbf {\bibinfo {volume} {94}},\
  \bibinfo {pages} {065010} (\bibinfo {year} {2016})},\ \Eprint
  {http://arxiv.org/abs/1605.06635} {arXiv:1605.06635 [gr-qc]} \BibitemShut
  {NoStop}%
\bibitem [{\citenamefont {Good}\ and\ \citenamefont
  {Ong}(2020)}]{good2020particle}%
  \BibitemOpen
  \bibfield  {author} {\bibinfo {author} {\bibfnamefont {M.~R.~R.}\
  \bibnamefont {Good}}\ and\ \bibinfo {author} {\bibfnamefont {Y.~C.}\
  \bibnamefont {Ong}},\ }\href {\doibase 10.1140/epjc/s10052-020-08761-7}
  {\bibfield  {journal} {\bibinfo  {journal} {Eur. Phys. J. C}\ }\textbf
  {\bibinfo {volume} {80}},\ \bibinfo {pages} {1169} (\bibinfo {year}
  {2020})},\ \Eprint {http://arxiv.org/abs/2004.03916} {arXiv:2004.03916
  [gr-qc]} \BibitemShut {NoStop}%
\bibitem [{\citenamefont {Good}\ \emph
  {et~al.}(2021{\natexlab{a}})\citenamefont {Good}, \citenamefont {Foo},\ and\
  \citenamefont {Linder}}]{Good:2020fjz}%
  \BibitemOpen
  \bibfield  {author} {\bibinfo {author} {\bibfnamefont {M.~R.~R.}\
  \bibnamefont {Good}}, \bibinfo {author} {\bibfnamefont {J.}~\bibnamefont
  {Foo}}, \ and\ \bibinfo {author} {\bibfnamefont {E.~V.}\ \bibnamefont
  {Linder}},\ }\href {\doibase 10.1088/1361-6382/abebb6} {\bibfield  {journal}
  {\bibinfo  {journal} {Class. Quant. Grav.}\ }\textbf {\bibinfo {volume}
  {38}},\ \bibinfo {pages} {085011} (\bibinfo {year} {2021}{\natexlab{a}})},\
  \Eprint {http://arxiv.org/abs/2006.01349} {arXiv:2006.01349 [gr-qc]}
  \BibitemShut {NoStop}%
\bibitem [{\citenamefont {Ford}\ and\ \citenamefont
  {Vilenkin}(1982)}]{Ford:1982ct}%
  \BibitemOpen
  \bibfield  {author} {\bibinfo {author} {\bibfnamefont {L.}~\bibnamefont
  {Ford}}\ and\ \bibinfo {author} {\bibfnamefont {A.}~\bibnamefont
  {Vilenkin}},\ }\href {\doibase 10.1103/PhysRevD.25.2569} {\bibfield
  {journal} {\bibinfo  {journal} {Phys. Rev. D}\ }\textbf {\bibinfo {volume}
  {25}},\ \bibinfo {pages} {2569} (\bibinfo {year} {1982})}\BibitemShut
  {NoStop}%
\bibitem [{\citenamefont {Reyes}(2021)}]{Reyes:2021npy}%
  \BibitemOpen
  \bibfield  {author} {\bibinfo {author} {\bibfnamefont {I.~A.}\ \bibnamefont
  {Reyes}},\ }\href {\doibase 10.1103/PhysRevLett.127.051602} {\bibfield
  {journal} {\bibinfo  {journal} {Phys. Rev. Lett.}\ }\textbf {\bibinfo
  {volume} {127}},\ \bibinfo {pages} {051602} (\bibinfo {year} {2021})},\
  \Eprint {http://arxiv.org/abs/2103.01230} {arXiv:2103.01230 [hep-th]}
  \BibitemShut {NoStop}%
\bibitem [{\citenamefont {Good}\ \emph
  {et~al.}(2020{\natexlab{a}})\citenamefont {Good}, \citenamefont {Linder},\
  and\ \citenamefont {Wilczek}}]{Good:2019tnf}%
  \BibitemOpen
  \bibfield  {author} {\bibinfo {author} {\bibfnamefont {M.~R.~R.}\
  \bibnamefont {Good}}, \bibinfo {author} {\bibfnamefont {E.~V.}\ \bibnamefont
  {Linder}}, \ and\ \bibinfo {author} {\bibfnamefont {F.}~\bibnamefont
  {Wilczek}},\ }\href {\doibase 10.1103/PhysRevD.101.025012} {\bibfield
  {journal} {\bibinfo  {journal} {Phys. Rev. D}\ }\textbf {\bibinfo {volume}
  {101}},\ \bibinfo {pages} {025012} (\bibinfo {year} {2020}{\natexlab{a}})},\
  \Eprint {http://arxiv.org/abs/1909.01129} {arXiv:1909.01129 [gr-qc]}
  \BibitemShut {NoStop}%
\bibitem [{\citenamefont {Moreno-Ruiz}\ and\ \citenamefont
  {Bermudez}(2022)}]{Moreno-Ruiz:2021qrf}%
  \BibitemOpen
  \bibfield  {author} {\bibinfo {author} {\bibfnamefont {A.}~\bibnamefont
  {Moreno-Ruiz}}\ and\ \bibinfo {author} {\bibfnamefont {D.}~\bibnamefont
  {Bermudez}},\ }\href {\doibase 10.1088/1361-6382/ac7506} {\bibfield
  {journal} {\bibinfo  {journal} {Class. Quant. Grav.}\ }\textbf {\bibinfo
  {volume} {39}},\ \bibinfo {pages} {145001} (\bibinfo {year} {2022})},\
  \Eprint {http://arxiv.org/abs/2112.00194} {arXiv:2112.00194 [gr-qc]}
  \BibitemShut {NoStop}%
\bibitem [{\citenamefont {Good}\ and\ \citenamefont
  {Linder}(2021)}]{Good:2020fsw}%
  \BibitemOpen
  \bibfield  {author} {\bibinfo {author} {\bibfnamefont {M.~R.~R.}\
  \bibnamefont {Good}}\ and\ \bibinfo {author} {\bibfnamefont {E.~V.}\
  \bibnamefont {Linder}},\ }\href {\doibase 10.1088/1367-2630/abe506}
  {\bibfield  {journal} {\bibinfo  {journal} {New J. Phys.}\ }\textbf {\bibinfo
  {volume} {23}},\ \bibinfo {pages} {043007} (\bibinfo {year} {2021})},\
  \Eprint {http://arxiv.org/abs/2003.01333} {arXiv:2003.01333 [gr-qc]}
  \BibitemShut {NoStop}%
\bibitem [{\citenamefont {Carlitz}\ and\ \citenamefont
  {Willey}(1987{\natexlab{b}})}]{CW2lifetime}%
  \BibitemOpen
  \bibfield  {author} {\bibinfo {author} {\bibfnamefont {R.~D.}\ \bibnamefont
  {Carlitz}}\ and\ \bibinfo {author} {\bibfnamefont {R.~S.}\ \bibnamefont
  {Willey}},\ }\href {\doibase 10.1103/PhysRevD.36.2336} {\bibfield  {journal}
  {\bibinfo  {journal} {Phys. Rev. D}\ }\textbf {\bibinfo {volume} {36}},\
  \bibinfo {pages} {2336} (\bibinfo {year} {1987}{\natexlab{b}})}\BibitemShut
  {NoStop}%
\bibitem [{\citenamefont {Good}\ and\ \citenamefont
  {Linder}(2022{\natexlab{a}})}]{Good:2021iny}%
  \BibitemOpen
  \bibfield  {author} {\bibinfo {author} {\bibfnamefont {M.~R.~R.}\
  \bibnamefont {Good}}\ and\ \bibinfo {author} {\bibfnamefont {E.~V.}\
  \bibnamefont {Linder}},\ }\href {\doibase 10.1088/1361-6382/ac60c3}
  {\bibfield  {journal} {\bibinfo  {journal} {Class. Quant. Grav.}\ }\textbf
  {\bibinfo {volume} {39}},\ \bibinfo {pages} {105003} (\bibinfo {year}
  {2022}{\natexlab{a}})},\ \Eprint {http://arxiv.org/abs/2108.07451}
  {arXiv:2108.07451 [gr-qc]} \BibitemShut {NoStop}%
\bibitem [{\citenamefont {Good}\ \emph
  {et~al.}(2021{\natexlab{b}})\citenamefont {Good}, \citenamefont {Mitra},\
  and\ \citenamefont {Zarikas}}]{2102.00158}%
  \BibitemOpen
  \bibfield  {author} {\bibinfo {author} {\bibfnamefont {M.~R.~R.}\
  \bibnamefont {Good}}, \bibinfo {author} {\bibfnamefont {A.}~\bibnamefont
  {Mitra}}, \ and\ \bibinfo {author} {\bibfnamefont {V.}~\bibnamefont
  {Zarikas}},\ }\href {\doibase 10.1134/S1063772921100115} {\bibfield
  {journal} {\bibinfo  {journal} {Astron. Rep.}\ }\textbf {\bibinfo {volume}
  {65}},\ \bibinfo {pages} {942} (\bibinfo {year} {2021}{\natexlab{b}})},\
  \Eprint {http://arxiv.org/abs/2102.00158} {arXiv:2102.00158 [gr-qc]}
  \BibitemShut {NoStop}%
\bibitem [{\citenamefont {Good}\ \emph
  {et~al.}(2020{\natexlab{b}})\citenamefont {Good}, \citenamefont {Zhakenuly},\
  and\ \citenamefont {Linder}}]{Good:2020byh}%
  \BibitemOpen
  \bibfield  {author} {\bibinfo {author} {\bibfnamefont {M.~R.~R.}\
  \bibnamefont {Good}}, \bibinfo {author} {\bibfnamefont {A.}~\bibnamefont
  {Zhakenuly}}, \ and\ \bibinfo {author} {\bibfnamefont {E.~V.}\ \bibnamefont
  {Linder}},\ }\href {\doibase 10.1103/PhysRevD.102.045020} {\bibfield
  {journal} {\bibinfo  {journal} {Phys. Rev. D}\ }\textbf {\bibinfo {volume}
  {102}},\ \bibinfo {pages} {045020} (\bibinfo {year} {2020}{\natexlab{b}})},\
  \Eprint {http://arxiv.org/abs/2005.03850} {arXiv:2005.03850 [gr-qc]}
  \BibitemShut {NoStop}%
\bibitem [{\citenamefont {Fern\'andez-Silvestre}\ \emph
  {et~al.}(2022)\citenamefont {Fern\'andez-Silvestre}, \citenamefont {Foo},\
  and\ \citenamefont {Good}}]{2109.04147}%
  \BibitemOpen
  \bibfield  {author} {\bibinfo {author} {\bibfnamefont {D.}~\bibnamefont
  {Fern\'andez-Silvestre}}, \bibinfo {author} {\bibfnamefont {J.}~\bibnamefont
  {Foo}}, \ and\ \bibinfo {author} {\bibfnamefont {M.~R.~R.}\ \bibnamefont
  {Good}},\ }\href {\doibase 10.1088/1361-6382/ac4b03} {\bibfield  {journal}
  {\bibinfo  {journal} {Class. Quant. Grav.}\ }\textbf {\bibinfo {volume}
  {39}},\ \bibinfo {pages} {055006} (\bibinfo {year} {2022})},\ \Eprint
  {http://arxiv.org/abs/2109.04147} {arXiv:2109.04147 [gr-qc]} \BibitemShut
  {NoStop}%
\bibitem [{\citenamefont {Good}\ and\ \citenamefont
  {Linder}(2018)}]{1711.09922}%
  \BibitemOpen
  \bibfield  {author} {\bibinfo {author} {\bibfnamefont {M.~R.~R.}\
  \bibnamefont {Good}}\ and\ \bibinfo {author} {\bibfnamefont {E.~V.}\
  \bibnamefont {Linder}},\ }\href {\doibase 10.1103/PhysRevD.97.065006}
  {\bibfield  {journal} {\bibinfo  {journal} {Phys. Rev. D}\ }\textbf {\bibinfo
  {volume} {97}},\ \bibinfo {pages} {065006} (\bibinfo {year} {2018})},\
  \Eprint {http://arxiv.org/abs/1711.09922} {arXiv:1711.09922 [gr-qc]}
  \BibitemShut {NoStop}%
\bibitem [{\citenamefont {Good}\ and\ \citenamefont
  {Linder}(2017)}]{Good:2017kjr}%
  \BibitemOpen
  \bibfield  {author} {\bibinfo {author} {\bibfnamefont {M.~R.~R.}\
  \bibnamefont {Good}}\ and\ \bibinfo {author} {\bibfnamefont {E.~V.}\
  \bibnamefont {Linder}},\ }\href {\doibase 10.1103/PhysRevD.96.125010}
  {\bibfield  {journal} {\bibinfo  {journal} {Phys. Rev. D}\ }\textbf {\bibinfo
  {volume} {96}},\ \bibinfo {pages} {125010} (\bibinfo {year} {2017})},\
  \Eprint {http://arxiv.org/abs/1707.03670} {arXiv:1707.03670 [gr-qc]}
  \BibitemShut {NoStop}%
\bibitem [{\citenamefont {DeWitt}(1975)}]{DeWitt:1975ys}%
  \BibitemOpen
  \bibfield  {author} {\bibinfo {author} {\bibfnamefont {B.~S.}\ \bibnamefont
  {DeWitt}},\ }\href {\doibase 10.1016/0370-1573(75)90051-4} {\bibfield
  {journal} {\bibinfo  {journal} {Phys. Rept.}\ }\textbf {\bibinfo {volume}
  {19}},\ \bibinfo {pages} {295} (\bibinfo {year} {1975})}\BibitemShut
  {NoStop}%
\bibitem [{\citenamefont {Ford}(1997)}]{Ford:1997hb}%
  \BibitemOpen
  \bibfield  {author} {\bibinfo {author} {\bibfnamefont {L.~H.}\ \bibnamefont
  {Ford}},\ }\href {https://arxiv.org/abs/gr-qc/9707062} {\emph {\bibinfo
  {title} {{Quantum field theory in curved space-time}}}}\ (\bibinfo
  {publisher} {9th Jorge Andre Swieca Summer School: Particles and Fields},\
  \bibinfo {year} {1997})\BibitemShut {NoStop}%
\bibitem [{\citenamefont {Fabbri}\ and\ \citenamefont
  {Navarro-Salas}(2005)}]{Fabbri}%
  \BibitemOpen
  \bibfield  {author} {\bibinfo {author} {\bibfnamefont {A.}~\bibnamefont
  {Fabbri}}\ and\ \bibinfo {author} {\bibfnamefont {J.}~\bibnamefont
  {Navarro-Salas}},\ }\href
  {https://www.worldscientific.com/doi/abs/10.1142/p378} {\emph {\bibinfo
  {title} {Modeling Black Hole Evaporation}}}\ (\bibinfo  {publisher} {Imperial
  College Press},\ \bibinfo {year} {2005})\BibitemShut {NoStop}%
\bibitem [{\citenamefont {Chen}\ and\ \citenamefont
  {Mourou}(2017)}]{Chen:2015bcg}%
  \BibitemOpen
  \bibfield  {author} {\bibinfo {author} {\bibfnamefont {P.}~\bibnamefont
  {Chen}}\ and\ \bibinfo {author} {\bibfnamefont {G.}~\bibnamefont {Mourou}},\
  }\href {\doibase 10.1103/PhysRevLett.118.045001} {\bibfield  {journal}
  {\bibinfo  {journal} {Phys. Rev. Lett.}\ }\textbf {\bibinfo {volume} {118}},\
  \bibinfo {pages} {045001} (\bibinfo {year} {2017})},\ \Eprint
  {http://arxiv.org/abs/1512.04064} {arXiv:1512.04064 [gr-qc]} \BibitemShut
  {NoStop}%
\bibitem [{\citenamefont {Akal}\ \emph {et~al.}(2021)\citenamefont {Akal},
  \citenamefont {Kusuki}, \citenamefont {Shiba}, \citenamefont {Takayanagi},\
  and\ \citenamefont {Wei}}]{Akal:2020twv}%
  \BibitemOpen
  \bibfield  {author} {\bibinfo {author} {\bibfnamefont {I.}~\bibnamefont
  {Akal}}, \bibinfo {author} {\bibfnamefont {Y.}~\bibnamefont {Kusuki}},
  \bibinfo {author} {\bibfnamefont {N.}~\bibnamefont {Shiba}}, \bibinfo
  {author} {\bibfnamefont {T.}~\bibnamefont {Takayanagi}}, \ and\ \bibinfo
  {author} {\bibfnamefont {Z.}~\bibnamefont {Wei}},\ }\href {\doibase
  10.1103/PhysRevLett.126.061604} {\bibfield  {journal} {\bibinfo  {journal}
  {Phys. Rev. Lett.}\ }\textbf {\bibinfo {volume} {126}},\ \bibinfo {pages}
  {061604} (\bibinfo {year} {2021})},\ \Eprint
  {http://arxiv.org/abs/2011.12005} {arXiv:2011.12005 [hep-th]} \BibitemShut
  {NoStop}%
\bibitem [{\citenamefont {Good}\ \emph
  {et~al.}(2021{\natexlab{c}})\citenamefont {Good}, \citenamefont {Lapponi},
  \citenamefont {Luongo},\ and\ \citenamefont {Mancini}}]{Good:2021asq}%
  \BibitemOpen
  \bibfield  {author} {\bibinfo {author} {\bibfnamefont {M.~R.~R.}\
  \bibnamefont {Good}}, \bibinfo {author} {\bibfnamefont {A.}~\bibnamefont
  {Lapponi}}, \bibinfo {author} {\bibfnamefont {O.}~\bibnamefont {Luongo}}, \
  and\ \bibinfo {author} {\bibfnamefont {S.}~\bibnamefont {Mancini}},\ }\href
  {\doibase 10.1103/PhysRevD.104.105020} {\bibfield  {journal} {\bibinfo
  {journal} {Phys. Rev. D}\ }\textbf {\bibinfo {volume} {104}},\ \bibinfo
  {pages} {105020} (\bibinfo {year} {2021}{\natexlab{c}})},\ \Eprint
  {http://arxiv.org/abs/2103.07374} {arXiv:2103.07374 [gr-qc]} \BibitemShut
  {NoStop}%
\bibitem [{\citenamefont {Sato}(2022)}]{Sato:2021ftf}%
  \BibitemOpen
  \bibfield  {author} {\bibinfo {author} {\bibfnamefont {Y.}~\bibnamefont
  {Sato}},\ }\href {\doibase 10.1103/PhysRevD.105.086016} {\bibfield  {journal}
  {\bibinfo  {journal} {Phys. Rev. D}\ }\textbf {\bibinfo {volume} {105}},\
  \bibinfo {pages} {086016} (\bibinfo {year} {2022})},\ \Eprint
  {http://arxiv.org/abs/2108.04637} {arXiv:2108.04637 [hep-th]} \BibitemShut
  {NoStop}%
\bibitem [{\citenamefont {Good}\ and\ \citenamefont
  {Abdikamalov}(2020)}]{Good:2020uff}%
  \BibitemOpen
  \bibfield  {author} {\bibinfo {author} {\bibfnamefont {M.~R.~R.}\
  \bibnamefont {Good}}\ and\ \bibinfo {author} {\bibfnamefont {E.}~\bibnamefont
  {Abdikamalov}},\ }\href {\doibase 10.3390/universe6090131} {\bibfield
  {journal} {\bibinfo  {journal} {Universe}\ }\textbf {\bibinfo {volume} {6}},\
  \bibinfo {pages} {131} (\bibinfo {year} {2020})},\ \Eprint
  {http://arxiv.org/abs/2008.08776} {arXiv:2008.08776 [gr-qc]} \BibitemShut
  {NoStop}%
\bibitem [{\citenamefont {Good}(2020)}]{good2020extreme}%
  \BibitemOpen
  \bibfield  {author} {\bibinfo {author} {\bibfnamefont {M.~R.~R.}\
  \bibnamefont {Good}},\ }\href {\doibase 10.1103/PhysRevD.101.104050}
  {\bibfield  {journal} {\bibinfo  {journal} {Phys. Rev. D}\ }\textbf {\bibinfo
  {volume} {101}},\ \bibinfo {pages} {104050} (\bibinfo {year} {2020})},\
  \Eprint {http://arxiv.org/abs/2003.07016} {arXiv:2003.07016 [gr-qc]}
  \BibitemShut {NoStop}%
\bibitem [{\citenamefont {Silva}\ \emph {et~al.}(2015)\citenamefont {Silva},
  \citenamefont {Braga}, \citenamefont {Rego},\ and\ \citenamefont
  {Alves}}]{Silva:2015yta}%
  \BibitemOpen
  \bibfield  {author} {\bibinfo {author} {\bibfnamefont {J.~D.~L.}\
  \bibnamefont {Silva}}, \bibinfo {author} {\bibfnamefont {A.~N.}\ \bibnamefont
  {Braga}}, \bibinfo {author} {\bibfnamefont {A.~L.~C.}\ \bibnamefont {Rego}},
  \ and\ \bibinfo {author} {\bibfnamefont {D.~T.}\ \bibnamefont {Alves}},\
  }\href {\doibase 10.1103/PhysRevD.92.025040} {\bibfield  {journal} {\bibinfo
  {journal} {Phys. Rev. D}\ }\textbf {\bibinfo {volume} {92}},\ \bibinfo
  {pages} {025040} (\bibinfo {year} {2015})}\BibitemShut {NoStop}%
\bibitem [{\citenamefont {Candelas}\ and\ \citenamefont
  {Deutsch}(1977)}]{Candelas:1977zza}%
  \BibitemOpen
  \bibfield  {author} {\bibinfo {author} {\bibfnamefont {P.}~\bibnamefont
  {Candelas}}\ and\ \bibinfo {author} {\bibfnamefont {D.}~\bibnamefont
  {Deutsch}},\ }\href {\doibase 10.1098/rspa.1977.0057} {\bibfield  {journal}
  {\bibinfo  {journal} {Proc. Roy. Soc. Lond. A}\ }\textbf {\bibinfo {volume}
  {354}},\ \bibinfo {pages} {79} (\bibinfo {year} {1977})}\BibitemShut
  {NoStop}%
\bibitem [{\citenamefont {Frolov}\ and\ \citenamefont
  {Serebriany}(1979)}]{Frolov_1979}%
  \BibitemOpen
  \bibfield  {author} {\bibinfo {author} {\bibfnamefont {V.~P.}\ \bibnamefont
  {Frolov}}\ and\ \bibinfo {author} {\bibfnamefont {E.~M.}\ \bibnamefont
  {Serebriany}},\ }\href {\doibase 10.1088/0305-4470/12/12/007} {\bibfield
  {journal} {\bibinfo  {journal} {Journal of Physics A: Mathematical and
  General}\ }\textbf {\bibinfo {volume} {12}},\ \bibinfo {pages} {2415}
  (\bibinfo {year} {1979})}\BibitemShut {NoStop}%
\bibitem [{\citenamefont {Frolov}\ and\ \citenamefont
  {Serebryany}(1980)}]{Frolov:1980pg}%
  \BibitemOpen
  \bibfield  {author} {\bibinfo {author} {\bibfnamefont {V.~P.}\ \bibnamefont
  {Frolov}}\ and\ \bibinfo {author} {\bibfnamefont {E.~M.}\ \bibnamefont
  {Serebryany}},\ }\href {\doibase 10.1088/0305-4470/13/10/017} {\bibfield
  {journal} {\bibinfo  {journal} {J. Phys. A}\ }\textbf {\bibinfo {volume}
  {13}},\ \bibinfo {pages} {3205} (\bibinfo {year} {1980})}\BibitemShut
  {NoStop}%
\bibitem [{\citenamefont {Good}\ and\ \citenamefont
  {Linder}(2022{\natexlab{b}})}]{Good:2021ffo}%
  \BibitemOpen
  \bibfield  {author} {\bibinfo {author} {\bibfnamefont {M.~R.~R.}\
  \bibnamefont {Good}}\ and\ \bibinfo {author} {\bibfnamefont {E.~V.}\
  \bibnamefont {Linder}},\ }\href {\doibase 10.1140/epjc/s10052-022-10167-6}
  {\bibfield  {journal} {\bibinfo  {journal} {Eur. Phys. J. C}\ }\textbf
  {\bibinfo {volume} {82}},\ \bibinfo {pages} {204} (\bibinfo {year}
  {2022}{\natexlab{b}})},\ \Eprint {http://arxiv.org/abs/2111.15148}
  {arXiv:2111.15148 [gr-qc]} \BibitemShut {NoStop}%
\bibitem [{\citenamefont {Zhakenuly}\ \emph {et~al.}(2021)\citenamefont
  {Zhakenuly}, \citenamefont {Temirkhan}, \citenamefont {Good},\ and\
  \citenamefont {Chen}}]{Zhakenuly:2021pfm}%
  \BibitemOpen
  \bibfield  {author} {\bibinfo {author} {\bibfnamefont {A.}~\bibnamefont
  {Zhakenuly}}, \bibinfo {author} {\bibfnamefont {M.}~\bibnamefont
  {Temirkhan}}, \bibinfo {author} {\bibfnamefont {M.~R.~R.}\ \bibnamefont
  {Good}}, \ and\ \bibinfo {author} {\bibfnamefont {P.}~\bibnamefont {Chen}},\
  }\href {\doibase 10.3390/sym13040653} {\bibfield  {journal} {\bibinfo
  {journal} {Symmetry}\ }\textbf {\bibinfo {volume} {13}},\ \bibinfo {pages}
  {653} (\bibinfo {year} {2021})},\ \Eprint {http://arxiv.org/abs/2101.02511}
  {arXiv:2101.02511 [gr-qc]} \BibitemShut {NoStop}%
\bibitem [{\citenamefont {Davies}(1982)}]{Davies:1982cn}%
  \BibitemOpen
  \bibfield  {author} {\bibinfo {author} {\bibfnamefont {P.~C.~W.}\
  \bibnamefont {Davies}},\ }\href {\doibase 10.1016/0370-2693(82)90824-3}
  {\bibfield  {journal} {\bibinfo  {journal} {Phys. Lett. B}\ }\textbf
  {\bibinfo {volume} {113}},\ \bibinfo {pages} {215} (\bibinfo {year}
  {1982})}\BibitemShut {NoStop}%
\bibitem [{\citenamefont {Walker}\ and\ \citenamefont
  {Davies}(1982)}]{Walker_1982}%
  \BibitemOpen
  \bibfield  {author} {\bibinfo {author} {\bibfnamefont {W.~R.}\ \bibnamefont
  {Walker}}\ and\ \bibinfo {author} {\bibfnamefont {P.~C.~W.}\ \bibnamefont
  {Davies}},\ }\href {\doibase 10.1088/0305-4470/15/9/008} {\bibfield
  {journal} {\bibinfo  {journal} {Journal of Physics A: Mathematical and
  General}\ }\textbf {\bibinfo {volume} {15}},\ \bibinfo {pages} {L477}
  (\bibinfo {year} {1982})}\BibitemShut {NoStop}%
\bibitem [{\citenamefont {Walker}(1985{\natexlab{a}})}]{Walker:1984ya}%
  \BibitemOpen
  \bibfield  {author} {\bibinfo {author} {\bibfnamefont {W.~R.}\ \bibnamefont
  {Walker}},\ }\href {\doibase 10.1088/0264-9381/2/2/006} {\bibfield  {journal}
  {\bibinfo  {journal} {Class. Quant. Grav.}\ }\textbf {\bibinfo {volume}
  {2}},\ \bibinfo {pages} {L37} (\bibinfo {year}
  {1985}{\natexlab{a}})}\BibitemShut {NoStop}%
\bibitem [{\citenamefont {Walker}(1985{\natexlab{b}})}]{walker1985particle}%
  \BibitemOpen
  \bibfield  {author} {\bibinfo {author} {\bibfnamefont {W.~R.}\ \bibnamefont
  {Walker}},\ }\href {\doibase 10.1103/PhysRevD.31.767} {\bibfield  {journal}
  {\bibinfo  {journal} {Phys. Rev. D}\ }\textbf {\bibinfo {volume} {31}},\
  \bibinfo {pages} {767} (\bibinfo {year} {1985}{\natexlab{b}})}\BibitemShut
  {NoStop}%
\bibitem [{\citenamefont {Ford}\ and\ \citenamefont
  {Roman}(2004)}]{Ford:2004ba}%
  \BibitemOpen
  \bibfield  {author} {\bibinfo {author} {\bibfnamefont {L.}~\bibnamefont
  {Ford}}\ and\ \bibinfo {author} {\bibfnamefont {T.~A.}\ \bibnamefont
  {Roman}},\ }\href {\doibase 10.1103/PhysRevD.70.125008} {\bibfield  {journal}
  {\bibinfo  {journal} {Phys. Rev. D}\ }\textbf {\bibinfo {volume} {70}},\
  \bibinfo {pages} {125008} (\bibinfo {year} {2004})},\ \Eprint
  {http://arxiv.org/abs/gr-qc/0409093} {arXiv:gr-qc/0409093} \BibitemShut
  {NoStop}%
\bibitem [{\citenamefont {Lynch}(2015)}]{Lynch:2015lra}%
  \BibitemOpen
  \bibfield  {author} {\bibinfo {author} {\bibfnamefont {M.~H.}\ \bibnamefont
  {Lynch}},\ }\href {\doibase 10.1103/PhysRevD.92.024019} {\bibfield  {journal}
  {\bibinfo  {journal} {Phys. Rev. D}\ }\textbf {\bibinfo {volume} {92}},\
  \bibinfo {pages} {024019} (\bibinfo {year} {2015})},\ \Eprint
  {http://arxiv.org/abs/1503.08891} {arXiv:1503.08891 [gr-qc]} \BibitemShut
  {NoStop}%
\bibitem [{\citenamefont {Akal}\ \emph {et~al.}(2022)\citenamefont {Akal},
  \citenamefont {Kawamoto}, \citenamefont {Ruan}, \citenamefont {Takayanagi},\
  and\ \citenamefont {Wei}}]{Akal:2022qei}%
  \BibitemOpen
  \bibfield  {author} {\bibinfo {author} {\bibfnamefont {I.}~\bibnamefont
  {Akal}}, \bibinfo {author} {\bibfnamefont {T.}~\bibnamefont {Kawamoto}},
  \bibinfo {author} {\bibfnamefont {S.-M.}\ \bibnamefont {Ruan}}, \bibinfo
  {author} {\bibfnamefont {T.}~\bibnamefont {Takayanagi}}, \ and\ \bibinfo
  {author} {\bibfnamefont {Z.}~\bibnamefont {Wei}},\ }\href@noop {} {\
  (\bibinfo {year} {2022})},\ \Eprint {http://arxiv.org/abs/2205.02663}
  {arXiv:2205.02663 [hep-th]} \BibitemShut {NoStop}%
\bibitem [{\citenamefont {Lynch}\ \emph {et~al.}(2021)\citenamefont {Lynch},
  \citenamefont {Cohen}, \citenamefont {Hadad},\ and\ \citenamefont
  {Kaminer}}]{Lynch:2019hmk}%
  \BibitemOpen
  \bibfield  {author} {\bibinfo {author} {\bibfnamefont {M.~H.}\ \bibnamefont
  {Lynch}}, \bibinfo {author} {\bibfnamefont {E.}~\bibnamefont {Cohen}},
  \bibinfo {author} {\bibfnamefont {Y.}~\bibnamefont {Hadad}}, \ and\ \bibinfo
  {author} {\bibfnamefont {I.}~\bibnamefont {Kaminer}},\ }\href {\doibase
  10.1103/PhysRevD.104.025015} {\bibfield  {journal} {\bibinfo  {journal}
  {Phys. Rev. D}\ }\textbf {\bibinfo {volume} {104}},\ \bibinfo {pages}
  {025015} (\bibinfo {year} {2021})},\ \Eprint
  {http://arxiv.org/abs/1903.00043} {arXiv:1903.00043 [gr-qc]} \BibitemShut
  {NoStop}%
\bibitem [{\citenamefont {Chen}\ \emph {et~al.}(2022)\citenamefont {Chen} \emph
  {et~al.}}]{AnaBHEL:2022sri}%
  \BibitemOpen
  \bibfield  {author} {\bibinfo {author} {\bibfnamefont {P.}~\bibnamefont
  {Chen}} \emph {et~al.} (\bibinfo {collaboration} {AnaBHEL}),\ }\href@noop {}
  {\  (\bibinfo {year} {2022})},\ \Eprint {http://arxiv.org/abs/2205.12195}
  {arXiv:2205.12195 [gr-qc]} \BibitemShut {NoStop}%
\bibitem [{\citenamefont {Byron}(1881)}]{Byron}%
  \BibitemOpen
  \bibfield  {author} {\bibinfo {author} {\bibfnamefont {G.~G.}\ \bibnamefont
  {Byron}},\ }in\ \href@noop {} {\emph {\bibinfo {booktitle} {Don Juan, Canto
  xv. Stanza 99.}}},\ \bibinfo {editor} {edited by\ \bibinfo {editor}
  {\bibfnamefont {M.}~\bibnamefont {Arnold}}}\ (\bibinfo  {publisher}
  {Macmillan and Co.},\ \bibinfo {address} {London},\ \bibinfo {year}
  {1881})\BibitemShut {NoStop}%
\bibitem [{\citenamefont {Sorge}(2018)}]{Sorge:2018zfd}%
  \BibitemOpen
  \bibfield  {author} {\bibinfo {author} {\bibfnamefont {F.}~\bibnamefont
  {Sorge}},\ }in\ \href@noop {} {\emph {\bibinfo {booktitle} {{15th Marcel
  Grossmann Meeting on Recent Developments in Theoretical and Experimental
  General Relativity, Astrophysics, and Relativistic Field Theories}}}}\
  (\bibinfo {year} {2018})\ \Eprint {http://arxiv.org/abs/1807.03968}
  {arXiv:1807.03968 [gr-qc]} \BibitemShut {NoStop}%
\bibitem [{\citenamefont {Sorge}\ and\ \citenamefont
  {Wilson}(2019)}]{Sorge:2019ecb}%
  \BibitemOpen
  \bibfield  {author} {\bibinfo {author} {\bibfnamefont {F.}~\bibnamefont
  {Sorge}}\ and\ \bibinfo {author} {\bibfnamefont {J.~H.}\ \bibnamefont
  {Wilson}},\ }\href {\doibase 10.1103/PhysRevD.100.105007} {\bibfield
  {journal} {\bibinfo  {journal} {Phys. Rev. D}\ }\textbf {\bibinfo {volume}
  {100}},\ \bibinfo {pages} {105007} (\bibinfo {year} {2019})},\ \Eprint
  {http://arxiv.org/abs/1909.07357} {arXiv:1909.07357 [gr-qc]} \BibitemShut
  {NoStop}%
\end{thebibliography}%

\end{document}